\documentclass[aps,prb,twocolumn,superscriptaddress]{revtex4-2}
\usepackage{graphicx}
\usepackage{subfigure}
\usepackage{dcolumn}
\usepackage{bm}
\usepackage{hyperref}
\usepackage{indentfirst}
\usepackage{braket}
\usepackage{amsmath}

\hypersetup{colorlinks=true, citecolor=blue, urlcolor=blue, linkcolor=blue}
\begin{document}
\title{Spin density wave and superconductivity in the bilayer $t$-$J$ model of $\rm{La}_{3}Ni_{2}O_{7}$ under renormalized mean-field theory}

\author{Yang Tian}
\affiliation{Department of Physics and State Key Laboratory of Surface Physics, Fudan University, Shanghai 200433, China}

\author{Yan Chen}
\email{yanchen99@fudan.edu.cn}
\affiliation{Department of Physics and State Key Laboratory of Surface Physics, Fudan University, Shanghai 200433, China}
\affiliation{Shanghai Branch, Hefei National Laboratory, Shanghai 201315, China}

\date{\today}

\begin{abstract}
Motivated by the recently discovered bilayer nickelate superconductor, the pressurized $\rm{La}_{3}Ni_{2}O_{7}$, we present a renormalized mean-field theory of a bilayer single-orbital $t$-$J$ model, highlighting the interplay between magnetism and superconductivity. We analyze the pairing symmetry and magnetic properties of the system, predicting two distinct states in which magnetism and superconductivity coexist. As hole doping increases, the magnetic order is rapidly suppressed. The inter-layer hopping \(t_{\perp}\) and coupling \(J_{\perp}\) promote a transition from intra-layer $d$-wave pairing to $s$-wave pairing, which is accompanied by a shift from antiferromagnetic order to a double spin stripe configuration. The latter has been extensively observed in ambient and high-pressure experiments. Our study offers theoretical insights into the coexistence of spin density waves and superconductivity.

\end{abstract}
\maketitle
\section{Introduction}

The discovery of unconventional superconductivity in the Ruddlesden-Popper series of nickel-based oxides, denoted as $\rm{R}_{n+1}Ni_{n}O_{3n+1}$ where $\rm{R}$ represents a rare earth element and $N$ indicates the number of consecutive layers, has attracted significant attention. Superconductivity has been observed at approximately 80 K in $\rm{La_{3}Ni_{2}O_{7}}$\cite{Nature621,Zhang2024,PhysRevLett.133.146002,PhysRevX.14.011040} and around 30 K in $\rm{La_{4}Ni_{3}O_{10}}$\cite{Nature631,doi:10.7566/JPSJ.93.053702,PhysRevB.109.144511,QingLi:17401,zhang2024superconductivitytrilayernickelatela4ni3o10} under high pressure. However, the mechanisms underlying superconductivity in these materials remain a subject of ongoing investigations. Additionally, the $\rm{La}_{3}Ni_{2}O_{7}$ may exhibit a more complex ordered phase because of the alternating hole distribution between $\rm{Ni}^{2+}$ and $\rm{Ni}^{3+}$, which may lead to a charge density wave (CDW) and spin density wave (SDW). Theoretical calculations suggest that SDW and CDW states may exist in $\rm{La}_3\rm{Ni}_2\rm{O}_7$ under ambient pressure\cite{doi:10.1021/ic960379j,Ni2025SpinDensityWave}, which originates from scattering between multiple Fermi surfaces. Random phase approximation (RPA) calculations predict SDW instabilities at wave vectors $(\pm \pi/2, \pm \pi/2)$\cite{PhysRevLett.131.126001} and diagonal stripes with period $2\pi/\delta$ at a small hole doping level $\delta$\cite{PhysRevLett.131.206501}. These density wave instabilities suggest the possible existence of more complex spin and charge ordered phases in the system.

Due to their similarities with cuprates, the interplay between superconductivity and magnetism is a crucial issue in the study of unconventional superconductors, particularly nickelates. Recent experiments provide a density-wave-like transition in $\rm{La}_{3}Ni_{2}O_{7-\delta}$ at ambient pressure\cite{PhysRevX.14.011040,Chen2024,xie20243221,ZHAO2025,gupta2024anisotropicspinstripedomains, PhysRevLett.132.256503,meng2024densitywavelikegapevolutionLa3Ni2O7}. The density wave is highly suppressed with increasing pressure, suggesting a competition with superconductivity. 
The resonant inelastic X-ray scattering (RIXS)\cite{Chen2024}, nuclear magnetic resonance (NMR)\cite{xie20243221,ZHAO2025}, resonant soft X-ray scattering (RSXS)\cite{gupta2024anisotropicspinstripedomains} and 
$\mu$sR\cite{PhysRevLett.132.256503,Khasanov2025} measurements suggest a SDW transition around $T_{SDW} \approx 150K$, with a vector of $(0.5\pi,0.5\pi)$. The magnetic ordering temperature increases by the pressure and approaches 155K at 2.31GPa\cite{Khasanov2025}. The ultrafast optical pump-probe spectroscopy measurements provide a density-wave-like order at high pressure\cite{meng2024densitywavelikegapevolutionLa3Ni2O7}. They proposed that the SDW observed at ambient pressure is gradually suppressed up to 13.3 GPa and completely vanishes around 26 GPa. Besides that, a distinct density-wave-like order emerges at pressures exceeding 29.4 GPa, with a transition temperature of approximately 135 K, likely associated with the predicted CDW order. 

There are many theoretical works for the bilayer nickelate $\rm{La}_{3}Ni_{2}O_{7}$\cite{PhysRevLett.131.126001,PhysRevB.108.L180510,chen2023criticalchargespininstabilities, PhysRevB.108.L201121, PhysRevLett.131.206501,gu2023, PhysRevB.108.L140505, PhysRevLett.131.236002, PhysRevB.108.L201108, PhysRevLett.132.106002, Shen_2023, PhysRevLett.132.036502,PhysRevB.109.L081105,PhysRevB.109.165154,lu2023superconductivitydopingsymmetricmass,PhysRevB.108.174501,PhysRevLett.132.126503,PhysRevB.108.214522,PhysRevLett.132.146002,PhysRevB.108.174511,PhysRevB.108.L140504,PhysRevB.108.125105,PhysRevB.109.144511,PhysRevB.110.L140508}. Most researchers believe that superconductivity primarily arises from the two $e_g$ orbitals ($d_{x^{2}-y^{2}}$ and $d_{3z^{2}-r^{2}}$). The $d_{x^{2}-y^{2}}$ orbital is nearly quarter-filled, while the $d_{3z^{2}-r^{2}}$ orbital is close to half-filled. The $d_{3z^{2}-r^{2}}$ orbitals are strongly coupled via inter-layer superexchange interactions mediated by the $O$-$2p_{z}$ orbital. Furthermore, these two $3d$ orbitals interact through onsite Hund’s coupling, which transfers the strong inter-layer antiferromagnetic (AFM) coupling from the $d_{3z^{2}-r^{2}}$ orbital to the $d_{x^{2}-y^{2}}$ orbital\cite{PhysRevLett.132.146002,PhysRevB.109.045154,PhysRevLett.132.036502}. Based on these facts, various multi-orbital and single-orbital models have been developed, predicting $s_\pm$-wave and $d$-wave pairing states. Apart from superconductivity, experimental investigations of density waves have revealed various spin- and charge-ordered states, which is also confirmed by several theoretical studies
\cite{qin2024intertwinedchargespindensity,zhang2024dopingevolutionnormalstate,Zhang2024stripe,labollita2024assessingformationspincharge,zhang2024emergentspinchargeorbitalordersuperconductor,lin2024magneticphasediagramtwoorbitaL,PhysRevB.110.L140508,PhysRevB.110.195135}. Some investigations argue that these states compete with superconductivity, suggesting that magnetic fluctuations could provide essential understanding for the pairing mechanisms behind high-temperature superconductivity in nickelates.

The linear spin density wave theory of a Heisenberg model with the third-nearest-neighbor AFM exchange coupling suggests several magnetic structures\cite{Chen2024,xie20243221}, where the effective inter-layer magnetic superexchange interaction is more substantial than the intra-layer magnetic interactions. The most likely spin configuration among the various magnetic structures, named the ``double spin stripe", is characterized by the ferromagnetic alignment of Ni atom spins in the $x$-$y$ direction and the alternating antiferromagnetic alignment in the $x$+$y$ direction. This stripe phase has been corroborated by other explorations employing various methods\cite{gupta2024anisotropicspinstripedomains,labollita2024assessingformationspincharge,zhang2024emergentspinchargeorbitalordersuperconductor}. Furthermore, certain studies assert the existence of spin stripes in the absence of significant charge disproportionation\cite{gupta2024anisotropicspinstripedomains}. However, this stripe phase was identified in the normal state of $\rm{La}_{3}Ni_{2}O_{7}$, and the interplay between the SDW and superconductivity remains unexplored.

We investigate a bilayer single-orbital $t$-$J$ model for $\rm{La}_{3}Ni_{2}O_{7}$ using renormalization mean-field theory (RMFT) to identify potential density wave orders and superconductivity. According to the previous theoretical works, the $3d_{z^2}$ and $3d_{x^2-y^2}$ electrons tend to form a spin-triplet state due to the strong Hund’s coupling ($J_{H} \gg J_{\perp}$) between two $e_{g}$ orbitals, and the inter-layer AFM superexchange interaction between the $3d_{z^2}$ spins will be transmitted to the $3d_{x^2-y^2}$ electrons\cite{PhysRevB.109.L081105}. Therefore, the two-orbital model can reduce to a single $3d_{x^2-y^2}$ orbital model by integrating out the $3d_{z^2}$ degrees of freedom, featuring an effective inter-layer AFM spin exchange between the two $3d_{x^2-y^2}$ orbitals situated in the two layers\cite{PhysRevLett.132.146002}. In addition, several recent studies have used single-orbital models to capture key aspects of nickelate superconductivity\cite{PhysRevLett.132.146002,PhysRevB.108.174511,PhysRevLett.132.036502,PhysRevB.110.024514,schlömer2024superconductivitypressurizednickelatela3ni2o7,lu2023superconductivitydopingsymmetricmass,PhysRevLett.133.126501}. Therefore, We believe that many of the essential features of $\rm{La}_{3}Ni_{2}O_{7}$, including the superconductivity and magnetism, can be qualitatively captured by such a single-orbital model. In the mean-field framework, we introduce four variational order parameters: the hole density $\delta_{li}$, the local spin moment representing antiferromagnetic correlations, the pair field $\Delta_{ij\sigma}$ indicating local electron pairing, and the bond order corresponding to the kinetic hopping term, where $i$ denotes a site position and $\langle ij\rangle$ represents the nearest-neighbor bond. At low doping, we identify a double spin stripe state characterized by the wave vector $Q = (\pi/2, \pi/2)$ and demonstrate that the inter-layer pairing of $d_{x^{2}-y^{2}}$ orbitals predominates and coexists with the SDW. Furthermore, the intra-layer pairing competes with the inter-layer pairing, which increases by the inter-layer magnetic coupling.

\section{Model Hamiltonian and Methods}
We start from the single-orbital $t$-$J$ model on a bilayer square lattice of the Ni-O plane, given by $H=H_{\parallel}+H_{\perp}$
\begin{equation}
\label{Hamiltonian}
\begin{split}
 H_{\parallel}=&-t\sum_{\langle ij\rangle, l,\sigma}(c_{li\sigma }^{\dagger }c_{lj\sigma}+h.c) +J_{1}\sum_{\langle ij\rangle,l}\mathbf{S}_{li}\cdot \mathbf{S}_{lj}\\
 &+J_{2}\sum_{\langle\langle\langle ij\rangle\rangle\rangle, l}\mathbf{S}_{li}\cdot \mathbf{S}_{lj},\\
 H_{\perp}=&-t_{\perp}\sum_{i,\sigma}(c_{1i\sigma }^{\dagger }c_{2i\sigma}+h.c)+J_{\perp}\sum_{i}\mathbf{S}_{1i}\cdot \mathbf{S}_{2i},
\end{split}  
\end{equation}
where $l=1,2$ labels the top and bottom layers, and $c_{li\sigma }^{\dagger}$ creates an electron at site $i$ with spin $\sigma$ on layer $l$. $S_{li}=\frac{1}{2}\sum_{\alpha\beta} c_{li\alpha }^{\dagger }\sigma_{\alpha\beta}c_{li\beta}$ is the spin operator. $t$ and $t_{\perp}$ is the hopping amplitude of intra-layer and inter-layer nearest-neighbor electrons, respectively.  The nearest-neighbor hopping $t$ is set to be the energy unit. The three magnetic exchange couplings $J_{1}$, $J_{2}$ and $J_{\perp}$ respectively represent the nearest-neighbor of intra-layer exchange interaction, third nearest-neighbor of the intra-layer exchange interaction, and nearest-neighbor of inter-layer exchange interaction.

The strong coupling constraint of no double occupancy is hard to study analytically. Zhang $et\:al$ introduced Gutzwiller renormalization factors to treat the constraint approximately\cite{PhysRevB.37.3759,article2}. Later, the antiferromagnetic (AFM) order case was considered by Himeda and Ogata\cite{article,PhysRevB.60.R9935}.  The resulting renormalized Hamiltonian is 
\begin{equation}
\begin{aligned}
 H^{\prime}=&H^{\prime}_{\parallel}+H^{\prime}_{\perp},\\
 H^{\prime}_{\parallel}=&-t\sum_{\langle ij\rangle, l,\sigma} g^{t}_{lij\sigma}(c_{li\sigma }^{\dagger}c_{lj\sigma}+h.c)\\
 &+J_{1}\sum_{\langle ij\rangle, l} (g^{s,z}_{lij}\mathbf{S}^{z}_{li}\mathbf{S}^{z}_{lj} +g^{s,xy}_{lij}(\frac{\mathbf{S}^{+}_{li}\mathbf{S}^{-}_{lj} + \mathbf{S}^{+}_{li}\mathbf{S}^{-}_{lj}}{2}))\\ 
 &+J_{2}\sum_{\langle\langle\langle ij\rangle\rangle\rangle, l}(g^{s,z}_{lij}\mathbf{S}^{z}_{li}\mathbf{S}^{z}_{lj} +g^{s,xy}_{lij}(\frac{\mathbf{S}^{+}_{li}\mathbf{S}^{-}_{lj} + \mathbf{S}^{+}_{li}\mathbf{S}^{-}_{lj}}{2})),\\
 H^{\prime}_{\perp}=&-t_{\perp}\sum_{i,\sigma} g^{t}_{ii\sigma\perp}(c_{1i\sigma }^{\dagger }c_{2i\sigma}+h.c)\\
 &+J_{\perp}\sum_{i}(g^{s,z}_{ii\perp}\mathbf{S}^{z}_{1i}\mathbf{S}^{z}_{2i} +g^{s,xy}_{ii\perp}(\frac{\mathbf{S}^{+}_{1i}\mathbf{S}^{-}_{2i} + \mathbf{S}^{+}_{1i}\mathbf{S}^{-}_{2i}}{2})).
\end{aligned}  
\end{equation}

The renormalization factors $g^{t}$ and $g^{s}$ used to evaluate a projected mean field wave function depend on the values of local magnetic, the corresponding pairing order parameters and hole density. The above local mean field parameters are defined as
\begin{equation}\label{meanfield}
\begin{aligned}
m^{\nu}_{li}&={\langle\Psi_{0}\mid S^{z}_{li}\mid\Psi_{0}\rangle},\\
\Delta^{\nu}_{lij\sigma}&=\sigma{\langle\Psi_{0}\mid c_{li\sigma }c_{lj\bar{\sigma}}\mid\Psi_{0}\rangle},\\
\chi^{\nu}_{lij\sigma}&={\langle\Psi_{0}\mid c_{li\sigma }^{\dagger}c_{lj\sigma}\mid\Psi_{0}\rangle},\\
\delta_{li}&= 1-{\langle\Psi_{0}\mid n_{li}\mid\Psi_{0}\rangle},\\
\end{aligned}
\end{equation}
where ${\mid\Psi_{0}\rangle}$ is the unprojected wave function. The superscript $\nu$ denotes that these quantities are unprojected and different from the real physical quantities. $\delta_{li}$, $\Delta^{\nu}_{lij\sigma}$ and $\chi^{\nu}_{lij\sigma}$ are hole density, pairing and hopping mean field on layer $l$, respectively. 

As for the Gutzwiller factors, we adopt the same form as used in previous studies\cite{Yang2009,PhysRevB.85.104511,Tu2016}

\begin{widetext}
\begin{equation}
\begin{aligned}
&g^{t}_{lij\sigma}=g^{t}_{li\sigma}g^{t}_{lj\sigma},\qquad\qquad\qquad\qquad\qquad
g^{t}_{li\sigma}=\sqrt{\frac{2\delta_{li}(1-\delta_{li})}{1-\delta_{li}^{2}+4(m^{\nu}_{li})^{2}}\frac{1+\delta_{li}+\sigma2m^{\nu}_{li}}{1+\delta_{li}-\sigma2m^{\nu}_{li}}},\\
&g^{s,xy}_{lij}=g^{s,xy}_{li}g^{s,xy}_{lj},\qquad\qquad\qquad\qquad\quad
g^{s,xy}_{li}=\frac{2\delta_{li}(1-\delta_{li})}{1-\delta_{li}^{2}+4(m^{\nu}_{li})^{2}},\\
g^{s,z}_{lij}=g^{s,xy}_{lij}&\frac{2((\bar{\Delta}^{\nu}_{lij})^{2}+(\bar{\chi}^{\nu}_{lij})^{2})-4m^{\nu}_{li}m^{\nu}_{lj}X^{2}_{ij} }{2((\bar{\Delta}^{\nu}_{lij})^{2}+(\overline{\chi}^{\nu}_{ij})^{2})-4m^{\nu}_{li}m^{\nu}_{lj}},\qquad
X_{ij}=1+\frac{12(1-\delta_{li})(1-\delta_{lj})((\bar{\Delta}^{\nu}_{lij})^{2}+(\bar{\chi}^{\nu}_{lij})^{2})}{\sqrt{(1-\delta_{li}^{2}+4(m^{\nu}_{li})^{2})(1-\delta_{lj}^{2}+4(m^{\nu}_{lj})^{2})}},
\end{aligned}
\end{equation}
where $\bar{\Delta}^{\nu}_{lij}=\sum_{l\sigma}\frac{\Delta^{\nu}_{lij\sigma}}{2}$ and $\bar{\chi}^{\nu}_{lij}=\sum_{l\sigma}\frac{\chi^{\nu}_{lij\sigma}}{2}$. 
The ground-state energy of the renormalized Hamiltonian can be obtained through the Gutzwiller renormalization factors and the mean-field order parameters
\begin{equation}\label{E}
\begin{aligned}
 E&={\langle\Psi_{0}\mid H^{\prime}\mid\Psi_{0}\rangle}\\
&=-t \sum_{\langle ij\rangle,\sigma}^{l=1,2}g^{t}_{lij\sigma}(\chi^{\nu}_{lij\sigma}+h.c)
-J_{1}\sum_{\langle ij\rangle,\sigma}^{l=1,2} ((\frac{g^{s,z}_{lij}}{4}+\frac{g^{s,xy}_{lij}}{2}\frac{\Delta^{\nu\ast}_{lij\bar{\sigma}}}{\Delta^{\nu\ast}_{lij\sigma}})\Delta^{\nu\ast}_{lij\sigma}\Delta^{\nu}_{lij\sigma}+(\frac{g^{s,z}_{lij}}{4}+\frac{g^{s,xy}_{lij}}{2}\frac{\chi^{\nu\ast}_{lij\bar{\sigma}}}{\chi^{\nu\ast}_{lij\sigma}})\chi^{\nu\ast}_{lij\sigma}\chi^{\nu}_{lij\sigma}) \\&-J_{2}\sum_{\langle\langle\langle ij\rangle\rangle\rangle,\sigma}^{l=1,2} ((\frac{g^{s,z}_{lij}}{4}+\frac{g^{s,xy}_{lij}}{2}\frac{\Delta^{\nu\ast}_{lij\bar{\sigma}}}{\Delta^{\nu\ast}_{lij\sigma}})\Delta^{\nu\ast}_{lij\sigma}\Delta^{\nu}_{lij\sigma} +(\frac{g^{s,z}_{lij}}{4}+\frac{g^{s,xy}_{lij}}{2}\frac{\chi^{\nu\ast}_{lij\bar{\sigma}}}{\chi^{\nu\ast}_{lij\sigma}})\chi^{\nu\ast}_{lij\sigma}\chi^{\nu}_{lij\sigma})  \\
 &-t_{\perp}\sum_{i,\sigma} g^{t}_{ii\sigma\perp}(\chi^{\nu}_{i\sigma\perp}+h.c)-J_{\perp}\sum_{i,\sigma}((\frac{g^{s,z}_{ii\perp}}{4}+\frac{g^{s,xy}_{lii\perp}}{2}\frac{\Delta^{\nu\ast}_{i\bar{\sigma}\perp}}{\Delta^{\nu\ast}_{i\sigma\perp}})\Delta^{\nu\ast}_{i\sigma\perp}\Delta^{\nu}_{i\sigma\perp}+(\frac{g^{s,z}_{ii\perp}}{4}+\frac{g^{s,xy}_{lii\perp}}{2}\frac{\chi^{\nu\ast}_{i\bar{\sigma}\perp}}{\chi^{\nu\ast}_{i\sigma\perp}})\chi^{\nu\ast}_{i\sigma\perp}\chi^{\nu}_{i\sigma\perp}) \\&+ J_{1}\sum_{\langle ij\rangle}^{l=1,2}g^{s,z}_{lij}m^{\nu}_{li}m^{\nu}_{lj} + J_{2}\sum_{\langle\langle\langle ij\rangle\rangle\rangle}^{l=1,2}g^{s,z}_{lij}m^{\nu}_{li}m^{\nu}_{lj} + J_{\perp}\sum_{i}g^{s,z}_{ii\perp}m^{\nu}_{1i}m^{\nu}_{2i}. 
\end{aligned}  
\end{equation}
\end{widetext}

Then we need to minimize the energy with two constraints:$\sum_{i}n_{i}=N_{e}$ and $\langle\Psi_{0}\mid \Psi_{0}\rangle=1$. We use the Lagrangian multiplier procedure to obtain the target function
\begin{equation}
\begin{split}
 W=&{\langle\Psi_{0}\mid H^{\prime}\mid\Psi_{0}\rangle} - \lambda(\langle\Psi_{0}\mid \Psi_{0}\rangle-1)\\&-\mu(\sum_{i}n_{i}-N_{e}).
\end{split}  
\end{equation}

The mean field Hamiltonian is given by $H_{MF}\mid \Psi_{0}\rangle = \lambda\mid \Psi_{0}\rangle$ and becomes
\begin{equation}\label{HMF}
\begin{split}
 H_{MF}=&\sum_{ij,l,\sigma}\frac{\partial W}{\partial \chi^{\nu}_{lij\sigma}}c^{\dagger}_{li\sigma}c_{lj\sigma}+h.c.\\&+\sum_{ij,l,\sigma}\frac{\partial W}{\partial \Delta^{\nu}_{lij\sigma}}\sigma c_{li\sigma}c_{lj\sigma}+h.c.\\&+\sum_{i,\sigma}\frac{\partial W}{\partial \chi^{\nu}_{i\sigma}}c^{\dagger}_{1i\sigma}c_{2i\sigma}+h.c.\\&+\sum_{i,\sigma}\frac{\partial W}{\partial \Delta^{\nu}_{i\sigma}}\sigma c_{1i\sigma}c_{2i\sigma}+h.c.\\&+\sum_{l,i,\sigma}\frac{\partial W}{\partial n_{li\sigma}}n_{li\sigma}.
\end{split}  
\end{equation}

The coefficients above are given as
\begin{equation}
\begin{split}
 \frac{\partial W}{\partial \chi^{\nu}_{lij\sigma}}=&-J_{p}(\frac{g^{s,z}_{lij}}{4}+\frac{g^{s,xy}_{lij}}{2}\frac{\chi^{\nu\ast}_{lij\bar{\sigma}}}{\chi^{\nu\ast}_{lij\sigma}})\chi^{\nu\ast}_{lij\sigma}-tg^{t}_{lij\sigma}+\frac{\partial W}{\partial g^{s,z}_{lij}}\frac{\partial g^{s,z}_{lij}}{\partial\chi^{\nu}_{lij\sigma}},\\
  \frac{\partial W}{\partial \Delta^{\nu}_{lij\sigma}}=&-J_{p}(\frac{g^{s,z}_{lij}}{4}+\frac{g^{s,xy}_{lij}}{2}\frac{\Delta^{\nu\ast}_{lij\bar{\sigma}}}{\Delta^{\nu\ast}_{lij\sigma}})\Delta^{\nu\ast}_{lij\sigma}+\frac{\partial W}{\partial g^{s,z}_{lij}}\frac{\partial g^{s,z}_{lij}}{\partial\Delta^{\nu}_{lij\sigma}},\\ 
  \frac{\partial W}{\partial \chi^{\nu}_{i\sigma}}=&-J_{\perp}(\frac{g^{s,z}_{12i}}{4}+\frac{g^{s,xy}_{12i}}{2}\frac{\chi^{\nu\ast}_{i\bar{\sigma}}}{\chi^{\nu\ast}_{i\sigma}})\chi^{\nu\ast}_{i\sigma}-t_{\perp}g^{t}_{i\sigma \perp}+\frac{\partial W}{\partial g^{s,z}_{12i}}\frac{\partial g^{s,z}_{12i}}{\partial\chi^{\nu}_{i\sigma}},\\
  \frac{\partial W}{\partial \Delta^{\nu}_{i\sigma}}=&-J_{\perp}(\frac{g^{s,z}_{12i}}{4}+\frac{g^{s,xy}_{12i}}{2}\frac{\Delta^{\nu\ast}_{i\bar{\sigma}}}{\Delta^{\nu\ast}_{i\sigma}})\Delta^{\nu\ast}_{i\sigma}+\frac{\partial W}{\partial g^{s,z}_{12i}}\frac{\partial g^{s,z}_{12i}}{\partial\Delta^{\nu}_{i\sigma}},\\  
  \frac{\partial W}{\partial n_{i\sigma}}=&-\mu - \frac{1}{2}\sigma\sum_{j}g^{s,z}_{lij}J_{p}m^{\nu}_{lj}-\sum_{j}\frac{\partial W}{\partial g^{s,xy}_{lij}}\frac{\partial g^{s,xy}_{lij}}{\partial n_{i\sigma}}\\&-\sum_{j}\frac{\partial W}{\partial g^{s,z}_{lij}}\frac{\partial g^{s,z}_{lij}}{\partial n_{i\sigma}}-\sum_{j\sigma\prime}\frac{\partial W}{\partial g^{t}_{lij\sigma\prime}}\frac{\partial g^{t}_{lij\sigma\prime}}{\partial n_{i\sigma}}.
\end{split}  
\end{equation}

Here, $J_{p}$($p=1,2$) corresponds to the nearest-neighbor and third-nearest-neighbor intra-layer antiferromagnetic couplings in Eq.\eqref{Hamiltonian}, respectively. This mean-field Hamiltonian $H_{MF}$ can be solved self-consistently; the eigenvalues and eigenvectors can determine the final mean-field parameters defined in Eq.\eqref{meanfield}.
After the self-consistency is achieved, we calculate the order parameters
\begin{equation}
\begin{split}
    \Delta_{\parallel,i}=&\sum_{\sigma}(g^{t}_{li,\sigma}g^{t}_{li+x,\bar{\sigma}}\Delta^{\nu}_{li,i+x,\sigma}+g^{t}_{li,\sigma}g^{t}_{li-x,\bar{\sigma}}\Delta^{\nu}_{li,i+x,\sigma}\\&+g^{t}_{li,\sigma}g^{t}_{li+y,\bar{\sigma}}\Delta^{\nu}_{li,i+y,\sigma}+g^{t}_{li,\sigma}g^{t}_{li-y,\bar{\sigma}}\Delta^{\nu}_{li,i-y,\sigma})/8,\\   \Delta_{\perp,i}=&\sum_{\sigma}g^{t}_{li,\sigma}g^{t}_{li+z,\bar{\sigma}}\Delta^{\nu}_{li,i+z,\sigma},\\
    m_{li}=&(\sqrt{g^{s,z}_{li,i+x}}+\sqrt{g^{s,z}_{li,i-x}}+\sqrt{g^{s,z}_{li,i+y}}+\sqrt{g^{s,z}_{li,i-y}},\\
    &+\sqrt{g^{s,z}_{li,i+2x}}+\sqrt{g^{s,z}_{li,i-2x}}+\sqrt{g^{s,z}_{li,i+2y}}+\sqrt{g^{s,z}_{li,i-2y}}\\
    &+\sqrt{g^{s,z}_{li,i+z}})m^{\nu}_{li}/9.\\
\end{split}
\end{equation}

\section{Numerical Results}

In the following, we fix the parameters at $t = 1, J_{1} = 0.3, J_{2} = 0.2$. According to the results of density functional theory (DFT) calculations, the hopping amplitude of $d_{x^{2}-y^{2}}$ orbital is estimated to be $t \sim 0.483 \, \text{eV}$\cite{PhysRevLett.131.126001,PhysRevB.108.L180510}, and we set $t = 1.0$ as the energy unit. First-principles calculations and dynamical mean-field theory (DMFT) calculations suggest the on-site Coulomb repulsion $U \sim 5 \, \text{eV}$\cite{PhysRevLett.131.206501,PhysRevB.109.L081105,Yang2024}. Using second-order perturbation theory in the strong-coupling limit, the effective magnetic interaction is given by $J = \frac{4t^2}{U}$. This yields a nearest-neighbor in-plane magnetic exchange $J_{1}$ in the range of 0.3 to 0.4,  while the inter-layer antiferromagnetic coupling is  stronger than the intra-layer antiferromagnetic interaction, approximately twice that of the intra-layer interaction\cite{PhysRevLett.132.036502,PhysRevLett.132.146002}.
The third-nearest-neighbor magnetic interaction $J_{2}$, derived from spin wave theory and supported by DFT calculations, is found to be comparable to $J_1$ \cite{Chen2024,xie20243221,Ni2025SpinDensityWave}. Given these considerations, we fix $J_{2} = 0.2$ in our model.

\begin{figure}
    \centerline{\includegraphics[width=\linewidth]{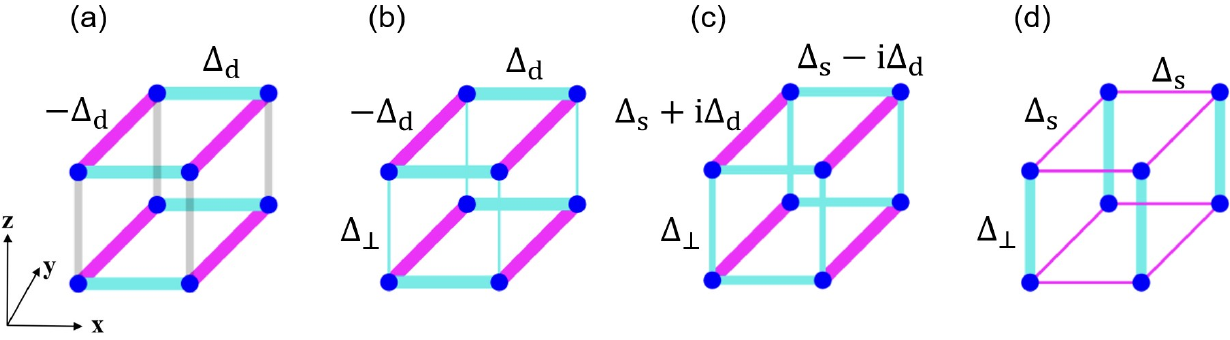}}
    \caption{\label{fig:fig0} The superconducting pairing configuration is shown, where blue (red) indicates that the superconducting order parameter is greater (less) than zero, and the black lines represent regions where the superconducting order parameter is zero. The thickness of the lines between lattice sites reflects the magnitude of the superconducting order parameter. (a) Intrinsic $d$-wave pairing; (b) Intrinsic $d$-wave pairing and inter-layer pairing; (c) Intrinsic $s+id$-wave pairing and inter-layer pairing; (d) Intrinsic $s$-wave pairing and inter-layer pairing.}
\end{figure}

We use an iterative method to self-consistently solve the equation \eqref{HMF}. Convergence is achieved when the change in the mean-field order parameter in equation \eqref{meanfield} between consecutive iterations is smaller than $10^{-3}$. All calculations are performed on an $8 \times 8$ square lattice. To obtain all possible magnetic orders, the transition term mean-field $\chi^{\nu}_{lij\sigma}$ is assumed to be uniform, while specific fluctuations are included in the other mean-fields. Despite this, we obtain several uniform superconducting pairing order parameters and two magnetic states. By tuning the inter-layer hopping ($t_\perp$) and the inter-layer antiferromagnetic exchange coupling ($J_\perp$), we identify four superconducting pairing configurations with different symmetries. As shown in Fig.\ref{fig:fig0}, these configurations include three types of intra-layer pairing: $d$-wave pairing, $s+id$-wave pairing, and $s$-wave pairing, along with inter-layer pairing. The inter-layer pairing order parameter $\Delta_{\perp}$ and the intra-layer $s$-wave pairing order parameter $\Delta^{s}_{\parallel}$ have opposite signs.

Similarly, for the magnetic order parameters, we find that the values of the order parameters for the upper and lower layers are equal in magnitude but opposite in sign, which is a direct consequence of the strong $J_\perp$ coupling. Hence, we can omit the layer index $l$.

Thus, we can define superconducting order parameters and magnetic order parameters as
\begin{equation}
\begin{split}
\Delta_{\parallel}&=\Delta_{\parallel,i}=\Delta^{s}_{\parallel}+i\Delta^{d}_{\parallel},\\
     \Delta_{\perp}&=\Delta_{\perp,i},\\
    m_{AF}&=\frac{1}{\sqrt{N}}\sum_{k}m_{i}e^{-i\pi R_{ix,iy}},\\
    m_{DS}&=\frac{1}{\sqrt{N}}\sum_{k}m_{i}e^{-i\frac{\pi}{2}R_{ix,iy}}.
\end{split}
\end{equation}
\subsection{Magnetism and superconductivity coexistence states}

\begin{figure}
\includegraphics[width=\linewidth]{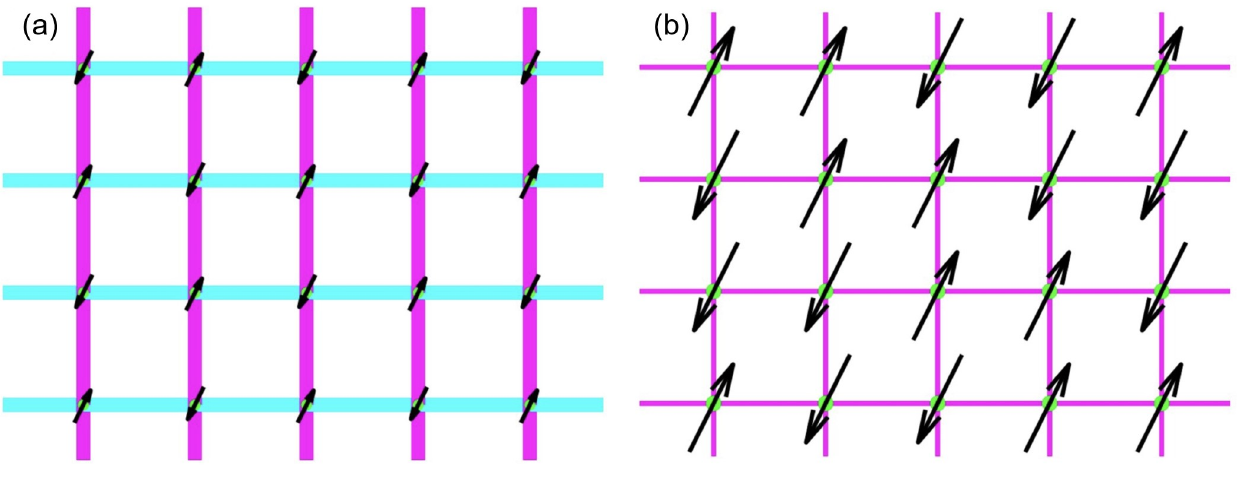}
\caption{\label{fig:fig1}\textbf Schematic illustration of modulations for stripe like patterns. (a) $d$-wave pairing and AFM with doping $\delta=0.1$ (b) $s$-wave and double spin stripe with doping $\delta=0.15$, respectively. The size of the green circle represents the hole density. The bond width around each site represents the amplitude of the superconducting order parameter, and the sign is positive (negative) for blue (red). The size of the black arrows represents the spin moment.}
\end{figure}

 We identify two types of magnetic ground states. In addition to the antiferromagnetism (AFM), we obtain double spin tripe with a period of four lattice spaces (4a$_0$), which consists of an up/up/down/down pattern of diagonal stripes. Both of them can coexist with superconductivity. Fig.\ref{fig:fig1} shows a schematic illustration of the pair field, charge density, and spin moment for the two magnetism and superconductivity coexistence states with a hole concentration of Fig.\ref{fig:fig1}(a) and 0.15 Fig.\ref{fig:fig1}(b). The magnitude of the pair field is proportional to the width of the bond; blue (red) denotes a positive (negative) value. The size of the arrow is proportional to the spin moment, and the size of the circle represents the hole density. Both charge and pairing are uniform. The AFM coexists with intra-layer $d$-wave pairing, and the double spin stripe coexists with $s$-wave pairing. 
 
 It is known that the coexistence of uniform $d$-wave pairing superconductivity and AFM is stable in the $t$-$J$ model. The third nearest neighbor coupling $J_{2}>0$ competes with $J_{1}$, leading to a double spin stripe state and breaking AFM simultaneously. The stripe is also inseparable from the bilayer structure, so it coexists with $s$-wave pairing and inter-layer pairing. Those two pairings also compete with $d$-wave pairing. 

\subsection{Doping effect}

Theoretically, the electron filling of the $\rm{Ni}$-$3d_{x^2-y^2}$ orbital in bilayer $\rm{La}_{3}Ni_{2}O_{7}$ should be equal to 0.5. However, due to hybridization with the $\rm{O}$-$2p$ orbitals and the $\rm{Ni}$-$3d_{z^2}$ orbital, there exists partial electron doping in the $\rm{Ni}$-$3d_{x^2-y^2}$ orbital. Under high pressure, the $\rm{Ni}$-$3d_{z^2}$ orbital crosses the Fermi level, generating holes, while an equal number of electrons transfer into the $\rm{Ni}$-$3d_{x^2-y^2}$ orbital. Consequently, the electron filling in the $\rm{Ni}$-$3d_{x^2-y^2}$ orbital further increases, and the corresponding energy band broadens\cite{Nature621}. Therefore, in the superconducting state, the electron filling number of the $3d_{x^2-y^2}$ orbital should be larger than 0.5. Besides, according to the previous RMFT calculations, the electron density of the $3d_{x^2-y^2}$ orbital in the pristine compound is determined to be $n=0.665$\cite{Luo2024}, and We focus on the electron-doped regime of the system. Hence, our calculations are performed within the electron filling range of $n = 0.65\sim0.9$ and the corresponding hole doping is from 0.1 to 0.35. 

\begin{figure}[t]
\centerline{\includegraphics[width=0.5\textwidth]{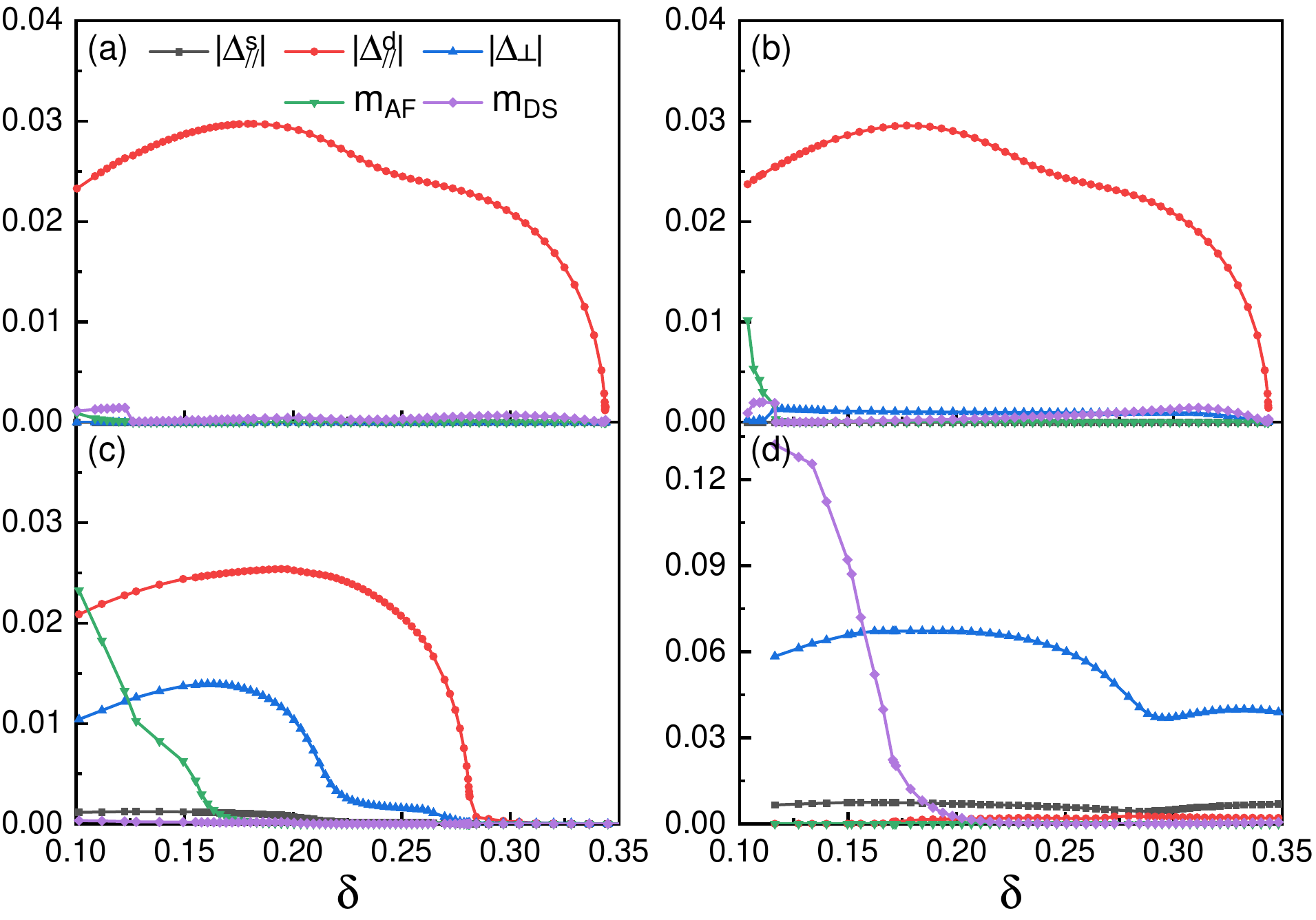}}
\caption{\label{fig:fig2} Superconducting pairing and magnetic order parameter as a function of hole doping $\delta$ for (a) $t_{\perp}=0,J_{\perp}=0$ (b) $t_{\perp}=0,J_{\perp}=0.3$ (c) $t_{\perp}=0.5,J_{\perp}=0.3$ (d) $t_{\perp}=0.5,J_{\perp}=0.5$. The intra-layer $s$-wave pairing exists only in nonzero $t_{\perp}$ and $J_{\perp}$. The double spin stripe mainly coexists with the inter-layer pairing.}
\end{figure}

Superconducting pairing and magnetic as a function of hole doping level with different $t_{\perp}$ and $J_{\perp}$ are displayed in Fig.\ref{fig:fig2}. At $t_{\perp}=0,J_{\perp}=0$, there is only intra-layer $d$-wave pairing $\Delta^{d}_{\parallel}$ because each layer is an individual $t$-$J$ model, see in Fig.\ref{fig:fig2}(a). Due to the existence of $J_{2}$, there is a small double spin stripe order, and the amplitude of this stripe is less than $10^{-5}$. Compared to the AFM order found in the $t$-$J$ model with only nearest neighbor magnetic interaction, $m_{AF}$ is much smaller. Still, the superconducting order parameter is more significant and exists in a broader range of $\delta$. This may be understood as the consequence of the nonzero value of $J_{2}$ in the present case, which competes with $J_{1}$, leading to a smaller AFM order.

The order parameters as a function of $\delta$ at $J_{\perp}=0.3$ are depicted in Fig.\ref{fig:fig2}(b). In this case, two layers are antiferromagnetically coupled by $J_{\perp}$. The intra-layer $d$-wave pairing $\Delta^{d}_{\parallel}$ is almost unchanged from that in Fig. \ref{fig:fig2}(a), and the inter-layer pairing $\Delta_{\perp}$ arises and stays around $10^{-3}$. The AFM order at low doping increases due to the contribution of the last term in Eq.\eqref{E}. The double spin stripe order is around $10^{-4}$, so we approximate that it does not coexist with the $d$-wave pairing.

In Fig.\ref{fig:fig2}(c), we display the order parameters as a function of $\delta$ at $t_{\perp}=0.5,J_{\perp}=0.3$. At finite doping, the intra-layer $s$-wave pairing $\Delta^{s}_{\parallel}$ appears, leading to an in-plane $s+id$-wave pairing with an inter-layer pairing which is also reported by Wu $et\:al$ \cite{PhysRevLett.132.146002}. The intra-layer $d$-wave pairing $\Delta^{d}_{\parallel}$ is suppressed but still the dominant pairing. The inter-layer pairing $\Delta_{\perp}$ increases to around $0.01$ with $t_{\perp}$ is up to $0.5$. This may explain the observed superconductivity in $\rm{La}_{3}Ni_{2}O_{7}$ under pressure. At low doping, the ground state has AFM order. As $\delta$ increases, the magnetism becomes a double spin stripe, and AFM vanishes. It is interesting to note that the intra-layer $s$-wave pairing $\Delta^{s}_{\parallel}$ survives only with nonzero values of $t_{\perp}$ and $J_{\perp}$, and  $\Delta^{s}_{\parallel}$ has opposite signs with inter-layer pairing $\Delta_{\perp}$. These may be explained by the fact that $\Delta^{s}_{\parallel}$ comes from the inter-layer pairing, which also can be interpreted by Eq.\eqref{DXZ}. 

We plot the the order parameters for $t_{\perp}=0.5,J_{\perp}=0.5$ in Fig.\ref{fig:fig2}(d). The intra-layer $d$-wave pairing $\Delta^{d}_{\parallel}$ is further suppressed at a larger $J_{\perp}$, the inter-layer pairing
$\Delta_{\perp}$ becomes the dominant pairing at the same time. $\Delta^{s}_{\parallel}$ increases as well, confirming the previous explanation. The inter-layer pairing $\Delta_{\perp}$ changes nonmonotonically concerning different doping levels, similar to doped two-leg spin-1/2 and spin-1 ladder\cite{article}. This may be the feature of bilayer or two-leg system\cite{PhysRevB.49.12058,Hou_2019}. As we can see, the double spin stripe increases rapidly, and AFM is completely suppressed. This gives rise to a coexistence state of intra-layer $s$-wave pairing $\Delta^{s}_{\parallel}$, inter-layer pairing $\Delta_{\perp}$ and double spin stripe.

In addition, Fig.\ref{fig:fig2} shows that magnetism is strongly suppressed with increasing doping, which is further confirmed by Fig.\ref{fig:fig3}(d). This behavior is consistent with our expectations and is also a generic feature of the $t$-$J$ model. Moreover, Lanczos calculations for the bilayer compound $\mathrm{La}_{3}\mathrm{Ni}_{2}\mathrm{O}_{7}$ reveal that spin-spin correlations decrease rapidly with hole doping, providing additional support for our findings \cite{PhysRevB.110.195135}.

\subsection{Analysis of pairing symmetry}

\begin{figure}
\includegraphics[width=\linewidth]{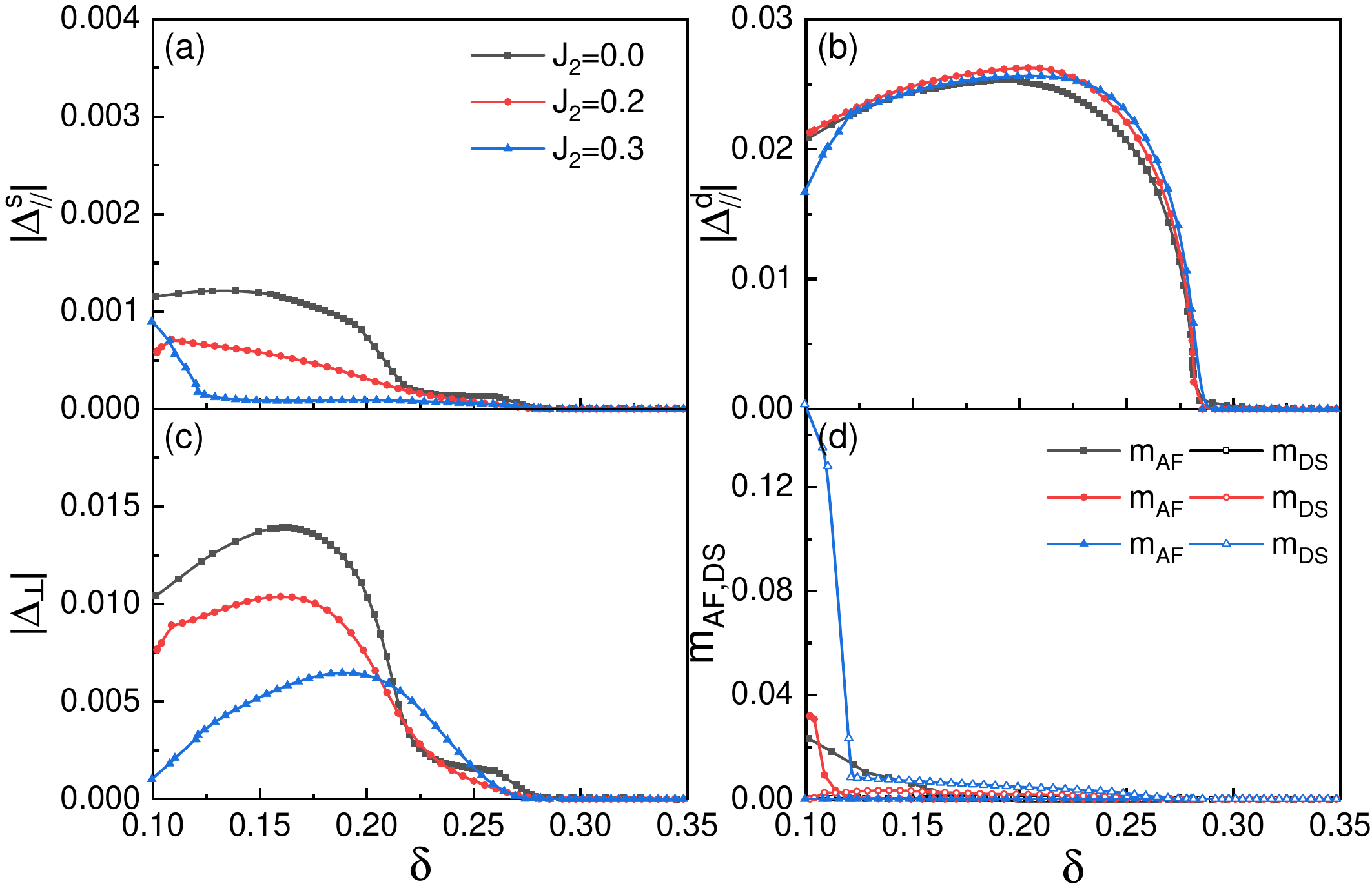}
\caption{\label{fig:fig3} Superconducting order parameter as a function of doping level with different $J_{2}$ at $t_{\perp}=0.5$, $J_{\perp}=0.3$. Superconducting order parameter for (a) intra-layer $s$-wave pairing, (b) intra-layer $d$-wave pairing, (c) inter-layer pairing, (d)AFM($m_{AF}$) and double spin stripe($m_{DS}$) respectively.}
\end{figure}

To give a further analysis of the problem, we switch to $k$ space to obtain some relations between the mean fields.
The mean-field Hamiltonian Eq.\eqref{HMF} can be written in momentum space as

\begin{equation}
 H_{k}=\sum_{k,\sigma}\epsilon_{k}c^{\dagger}_{k,\sigma }c_{k,\sigma}+h.c.+\sum_{k}\Tilde{\Delta}_{k}c^{\dagger}_{k\uparrow}c^{\dagger}_{-k\downarrow}+h.c.,
\end{equation}
where
\begin{equation}
\begin{split}
\epsilon_{k_{1}}=&-2g^{t}t(\cos{k_{x}}+\cos{k_{y}})-g^{t}t_{\perp}\cos{k_{z}}\\
&-\frac{3}{4}g^{s}J_{1}(2\chi_{x}\cos{k_{x}}+2\chi_{y}\cos{k_{y}})\\
&-\frac{3}{4}g^{s}J_{2}(2\chi^{\prime}_{x}\cos{2k_{x}}+2\chi^{\prime}_{y}\cos{2k_{y}})\\
&-\frac{3}{4}g^{s}J_{\perp}\chi_{z}\cos{k_{z}},
\end{split}
\end{equation}
\begin{equation}
\begin{split}
\Tilde{\Delta}_{k_{1}}=&-\frac{3}{4}g^{s}J_{1}(2\Delta_{x}\cos{k_{x}}+2\chi_{y}\cos{k_{y}})\\
&-\frac{3}{4}g^{s}J_{2}(2\Delta^{\prime}_{x}\cos{2k_{x}}+2\Delta^{\prime}_{y}\cos{2k_{y}})\\
&-\frac{3}{4}g^{s}J_{\perp}\Delta_{z}\cos{k_{z}},
\end{split}
\end{equation}

\begin{equation}
    \epsilon_{k}=2\epsilon_{k_{1}},\qquad \Tilde\Delta_{k}=2\Tilde\Delta_{k_{1}}.
\end{equation}

$\epsilon_{k_{1}}$ and $\tilde{\Delta}_{k_{1}}$ include both the intra-layer energy of layer 1 and the energy of inter-layer hopping and magnetic interaction with layer 2. The mean-field energy of layer 2 is equal to those of layer 1. Within the range of values in this article, $J_{2}$ does not affect the pairing symmetry, see in Fig.\ref{fig:fig3}, and neither does magnetism. Thus we can set $J_{2}=0$, $J_{\perp}=J_{1}$ to analyze the pairing symmetry. We consider the half-filled case with $g^{t}=0$. The total energy of the system is
 \begin{equation}
     E=-\frac{3}{4}g^{s}J_{1}\sum_{k}E_{k},
 \end{equation}
 where
 \begin{equation}\label{ek}
     E_{k}=\sqrt{\chi^{2}_{k}+\Delta^{2}_{k}},
 \end{equation}
\begin{equation}\label{xk}
    \chi_{k}=2(\chi_{x}\cos{k_{x}}+\chi_{y}\cos{k_{y}})+\chi_{z}\cos{k_{z}},
\end{equation}
\begin{equation}\label{dk}
\Delta_{k}=2(\Delta_{x}\cos{k_{x}}+\Delta_{y}\cos{k_{y}})+\Delta_{z}\cos{k_{z}}.
\end{equation}

Following the method with \cite{PhysRevB.37.3759,PhysRevB.49.12058,PhysRevB.99.094510}, we use an ansatz for $E_{k}$
\begin{equation}
    E_{k}=C\sqrt{\cos^{2}{k_{x}}+\cos^{2}{k_{y}}+\cos^{2}{k_{z}}},
\end{equation}
where C is a parameter to be determined. The reason for choosing this ansatz is that the hoppings and interactions between lattice sites in the Hamiltonian are along the $x$, $y$, and $z$ directions. Therefore, when transforming to momentum space, the energy should include terms like $\cos{k_{x}}$, $\cos{k_{y}}$, and $\cos{k_{z}}$. Note that there are only two sites in the $z$ direction, so $k_{z} = -\pi, \pi$. Under this constraint, combining Eq.\eqref{ek}, \eqref{xk}, and \eqref{dk}, we can derive the $s+id$-wave pairing in the $xy$ plane, which satisfies

\begin{equation}
\begin{gathered}
    \Delta_{x}=\Delta_{s,x}+i\Delta_{d,x},\\
    \Delta_{y}=\Delta_{s,y}+i\Delta_{d,y}=\Delta_{s,x}-i\Delta_{d,x},\\
    \Delta^{2}_{d}-\Delta^{2}_{s}=\chi^{2}_{x}=\chi^{2}_{y}.
\end{gathered}
\end{equation}

The inter-layer mean fields should satisfy
\begin{equation}\label{DXZ}
    \begin{gathered}
        2\chi_{x}\chi_{z}+(\Delta_{z}\Delta^{\star}_{x}+\Delta_{x}\Delta^{\star}_{z})=0,\\
        2\chi_{y}\chi_{z}+(\Delta_{z}\Delta^{\star}_{y}+\Delta_{y}\Delta^{\star}_{z})=0,\\
        \chi_{x}\chi_{z}+\Delta_{z}\Delta_{s}=0.
    \end{gathered}
\end{equation}

The mean fields satisfy the following simultaneous equations:
\begin{equation}\label{constraints}
\begin{gathered}
    \chi^{2}_{1}+\Delta^{2}_{1}=\chi^{2}_{z}+\Delta^{2}_{z}=C^{2},\\
    \Delta^{2}_{d}-\Delta^{2}_{s}=\chi^{2}_{1},\\
    \chi_{1}\Delta_{1}=\chi_{z}\Delta_{s},
\end{gathered}   
\end{equation}
where 
\begin{equation}
\begin{gathered}
    \chi_{1}=\chi_{x}=\chi_{y},\\
    \Delta_{1}=\Delta_{s}+i\Delta_{d}.
\end{gathered}
\end{equation}

\begin{figure}
\includegraphics[width=\linewidth]{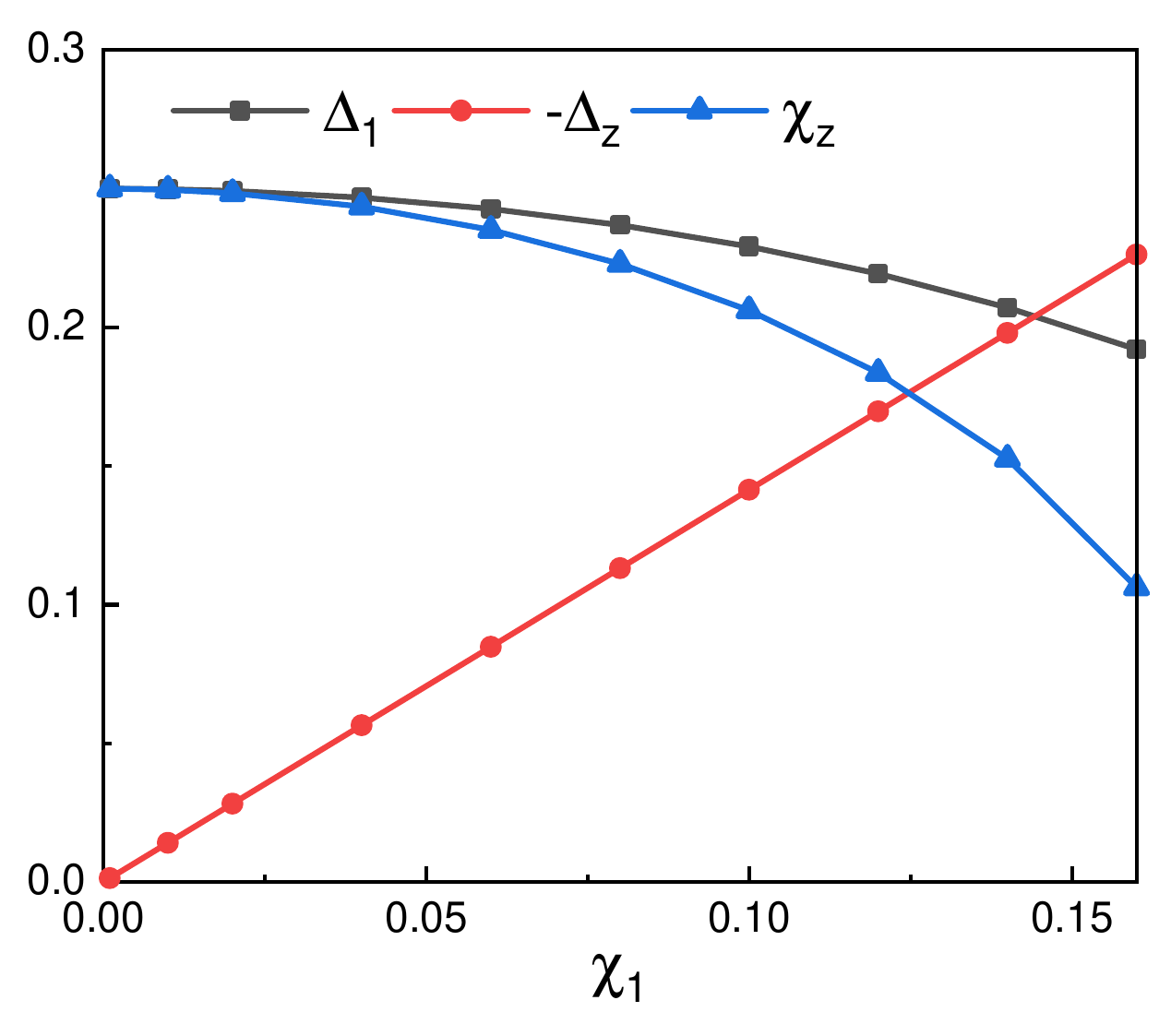}
\caption{\label{fig:fig4} The mean field $\Delta_{1},\Delta{z},\chi_{z}$ as a function of $\chi_{1}$ at half filling. Each point is a degenerate state and satisfies the constraints Eq.\eqref{constraints}.}
\end{figure}

According to the third equation in \eqref{DXZ}, it is clear that the inter-layer pairing mean field $\Delta_{z}$ and the intra-layer $s$-wave pairing mean field $\Delta_{s}$ have opposite signs. This is consistent with the results of the RMFT calculations shown in Fig.\ref{fig:fig0}.

We remark that there is only one independent field for fixed C. For example, if we fix $\chi_{1}$, the absolute values of all mean fields are determined. We can obtain many degenerate states by changing $\chi_{1}$. To verify this result, we solve the mean-field Hamiltonian Eq.\eqref{HMF} at half filling and obtain some states with the same energy as shown in Fig.\ref{fig:fig4}. The inter-layer pairing field $\Delta_{z}$ increases with the increasing of $\chi_{1}$, while intra-layer pairing field $\Delta_{1}$ and inter-layer $\chi_{z}$ field decrease. Each point is a degenerate state, which satisfies the constraints Eq.\eqref{constraints}. 

\begin{figure}
\includegraphics[width=\linewidth]{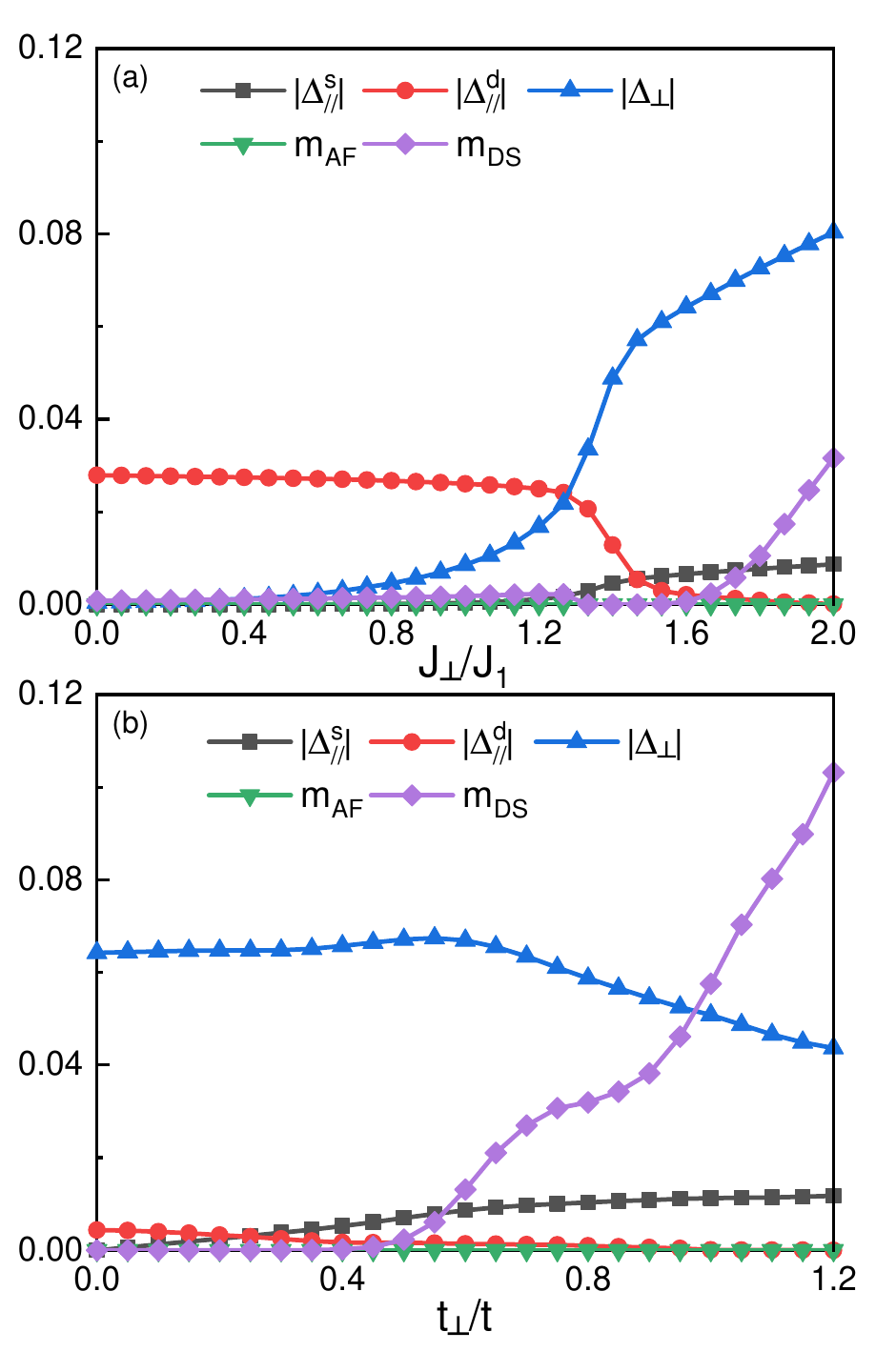}
\caption{\label{fig:fig5} (a) Superconducting order parameter and magnetic order parameter as a function of $J_{\perp}/J_{1}$ with doping $\delta=0.2$ for $t_{\perp}=0.5$. (b) Superconducting order parameter and magnetic order parameter as a function of $t_{\perp}/t_{1}$ with doping $\delta=0.2$ for $J_{\perp}=0.5$.}
\end{figure}

When slightly doping holes, the above constraints are still valid. If we fix the intra-layer hopping $t$ and raise the inter-layer hopping $t_{\perp}$ from zero, $\chi_{z}$ increase to enhance the absolute value of inter-layer kinetic momentum, but $\chi_{1}$ decreases at the same time, which will lower intra-layer kinetic momentum and leads to a smaller absolute value of total kinetic momentum. Because $t_{\perp }$ is small from the beginning, the contribution of the inter-layer kinetic energy is small, and the $\chi_{1}$ is dominant. There is no obvious change at small $t_{\perp}$. When $t_{\perp}$ keeps growing, $\chi_{z}$ will increase gradually and $\Delta_{z}$ decrease, $\Delta_{1}$(actually is $\Delta_{s}$) increases. The RMFT results in Fig.\ref{fig:fig5}(b) also support the conclusion of momentum-space analysis.

\subsection{Effects of $J_{\perp}$ and $t_{\perp}$}

In this section, we briefly discuss the effects of inter-layer hopping and coupling. According to our analysis before, the inter-layer pairing strength would increase with $J_{\perp}/J_{1}$ because it directly comes from the mean-field decoupling of the antiferromagnetic interaction $J_{\perp}$ as shown in Fig.\ref{fig:fig5}(a). The intra-layer $d$-wave pairing $\Delta^{d}_{\parallel}$ decreases with the increasing of $J_{\perp}/J_{1}$ due to the competition with the inter-layer pairing $\Delta_{\perp}$. The electron at site $i$ chooses NN intra-layer or inter-layer electrons to form a Cooper pair. The inter-layer $s$-wave pairing comes from the common effect between $t_{\perp}$ and $J_{\perp}$. It competes with intra-layer $d$-wave pairing and increases monotonically with $J_{\perp}/J_{1}$. The bilayer structure is significant to the double spin stripe. Although it is an in-plane stripe, the corresponding magnetic order parameter $m_{DS}$ increases with $J_{\perp}/J_{1}$ and $t_{\perp}/t$.The inter-layer pairing $\Delta_{z}$ barely changes with $t_{\perp}/t$ in the  $0<t_{\perp}/t<0.55$ and obviously decreases when $t_{\perp}$ continue to increase. The intra-layer $d$-wave pairing and $s$-wave pairing monotonously decrease and increase with $t_{\perp}/t$, respectively. Those RMFT results are consistent with the conclusion of momentum-space analysis. 

Previous studies have demonstrated that strong inter-layer antiferromagnetic coupling induces hole pairing and the emergence of a spin gap upon doping \cite{PhysRevB.45.5744}. Moreover, both the spin gap and the hole-pair binding energy increase with the strength of the inter-layer coupling, suggesting an enhancement of superconducting correlations. These findings agree with our RMFT results, further supporting the role of inter-layer interactions in promoting superconductivity.

\section{Conclusion}
In this work, we investigate the interplay between superconductivity and magnetism within a single-orbital bilayer $t$-$J$ model, proposing an intra-layer $s+id$-wave pairing and a dominant inter-layer pairing. The analysis of pairing symmetry indicates a unified solution with numerical results. The intra-layer $d$-wave pairing coexists with antiferromagnetism (AFM), and the intra-layer $s$-wave pairing coexists with a double spin stripe. The transition between these states is tuned by inter-layer hopping $t_{\perp}$ and inter-layer coupling $J_{\perp}$. 

Compared with previous theoretical studies, we consider the magnetism and identify the double spin stripe, which is anticipated by experiments. Furthermore, we propose a coexistence state of superconductivity and the double spin stripe, highlighting its evolution with $t_{\perp}$ and $J_{\perp}$, which significantly increase under pressure. The variation in the amplitude of superconductivity with changes in inter-layer parameters aligns with experimental findings. We would like to emphasize that we aim to provide qualitative insights into how doping and inter-layer interactions affect the pairing symmetry and magnetic ordering in a bilayer system. We believe that the larger electron occupancy in the $\rm{Ni}$-$3d_{x^2-y^2}$ orbital provides a better description of the $\rm{La}_{3}Ni_{2}O_{7}$ under high pressure.

There are some predictions about the double spin stripe in bilayer $\rm{La}_{3}Ni_{2}O_{7}$ at ambient pressure\cite{Chen2024,PhysRevLett.132.256503,gupta2024anisotropicspinstripedomains,labollita2024assessingformationspincharge,zhang2024emergentspinchargeorbitalordersuperconductor}, the density wave is suppressed by pressure which can enhance superconductivity. But a result of $\mu$sR shows that the magnetic ordering temperature was observed to rise under increasing pressure\cite{Khasanov2025}. Recently, an ultrafast optical pump-probe spectroscopy experiment reported a pressure-induced separation between two density wave-like orders\cite{meng2024densitywavelikegapevolutionLa3Ni2O7}, which is in agreement with the first-principles calculations\cite{PhysRevB.110.L140508}. They found the magnetic order aligns with the critical temperature associated with the density wave anomaly. The above two experiments confirm our numerical results in some ways. However, some problems remain to be solved: (a) Our findings demonstrate that pressure enhances the spin density wave and superconductivity in their coexisting region. However, experimental evidence indicates that the enhancement of pressure on magnetic order is confined to a narrow range around ambient pressure, and superconductivity does not appear\cite{Khasanov2025}. (b)We have not considered the effects of the interplay between the $d_{x^{2}-y^{2}}$ orbital and the $d_{z^{2}}$ orbital. In particular, we recognize that the $d_{z^2}$ orbital may cross the Fermi level in the superconducting state, potentially creating a complex Fermi surface and momentum-dependent inter-layer hopping, such as $(\cos k_x - \cos k_y)^2$, as discussed in Ref.\cite{PhysRevLett.131.126001,cpl417077402}, an effect absent in our framework. Moreover, we believe that the absence of the CDW may be related to the single orbital. Because the alternating charge distribution between $\rm{Ni}^{2+}$ and $\rm{Ni}^{3+}$ ions involves in both $eg$ orbitals. Specifically, $\rm{Ni}^{2+}$ sites are characterized by one hole residing in the $d_{x^{2}-y^{2}}$ orbital and another in the $d_{z^{2}}$ orbital, whereas $\rm{Ni}^{3+}$ sites correspond to two holes in the $d_{x^{2}-y^{2}}$ orbital and the third one in the $d_{z^{2}}$ orbital. Whether CDW exists in the superconducting state needs further investigation. 

However, while some experiments and theoretical works suggest the involvement of the $d_{z^2}$ orbital near the Fermi level under pressure, strong electron correlations may still render it nearly localized. For instance, DMFT calculations indicate that the local spin correlation of the $d_{z^2}$ orbital is significantly stronger than that of the $d_{x^2-y^2}$ orbital\cite{Wu2024}, implying that $d_{z^2}$ electrons may contribute mainly to local moment formation rather than itinerant states. 
Hence, in the coexistence regime, we treat $d_{z^2}$ electrons as quasi-localized, providing magnetic moments without participating significantly in conduction.

Our model captures key qualitative aspects, such as the coexistence of superconductivity and spin density wave and their evolution under pressure, but it is still based on a simplified single-orbital framework. The complex multi-orbital nature of $\rm{La}_{3}Ni_{2}O_{7}$, including possible self-doping effects between $e_g$ orbitals\cite{PhysRevB.108.L201108,PhysRevLett.132.106002,Shen_2023,PhysRevLett.132.036502,PhysRevB.109.L081105}, lies beyond the current scope of this work. These additional effects may significantly influence orbital occupations and the emergence or suppression of CDW and superconductivity. Nevertheless, our study provides theoretical support for SDW-superconductivity coexistence and establishes a foundation for understanding their interplay in pressurized $\rm{La}_{3}Ni_{2}O_{7}$.

\onecolumngrid
\section*{Acknowlegements}
We are grateful to Ting-Kuo Lee for many useful discussions. This work is supported by the National Key Research and Development Program of China Grant No. 2022YFA1404204, and the National Natural Science Foundation of China Grant No. 12274086.
\nocite{*}

\twocolumngrid 
\bibliography{327sdw}

\begin{thebibliography}{67}%
\makeatletter
\providecommand \@ifxundefined [1]{%
 \@ifx{#1\undefined}
}%
\providecommand \@ifnum [1]{%
 \ifnum #1\expandafter \@firstoftwo
 \else \expandafter \@secondoftwo
 \fi
}%
\providecommand \@ifx [1]{%
 \ifx #1\expandafter \@firstoftwo
 \else \expandafter \@secondoftwo
 \fi
}%
\providecommand \natexlab [1]{#1}%
\providecommand \enquote  [1]{``#1''}%
\providecommand \bibnamefont  [1]{#1}%
\providecommand \bibfnamefont [1]{#1}%
\providecommand \citenamefont [1]{#1}%
\providecommand \href@noop [0]{\@secondoftwo}%
\providecommand \href [0]{\begingroup \@sanitize@url \@href}%
\providecommand \@href[1]{\@@startlink{#1}\@@href}%
\providecommand \@@href[1]{\endgroup#1\@@endlink}%
\providecommand \@sanitize@url [0]{\catcode `\\12\catcode `\$12\catcode `\&12\catcode `\#12\catcode `\^12\catcode `\_12\catcode `\%12\relax}%
\providecommand \@@startlink[1]{}%
\providecommand \@@endlink[0]{}%
\providecommand \url  [0]{\begingroup\@sanitize@url \@url }%
\providecommand \@url [1]{\endgroup\@href {#1}{\urlprefix }}%
\providecommand \urlprefix  [0]{URL }%
\providecommand \Eprint [0]{\href }%
\providecommand \doibase [0]{https://doi.org/}%
\providecommand \selectlanguage [0]{\@gobble}%
\providecommand \bibinfo  [0]{\@secondoftwo}%
\providecommand \bibfield  [0]{\@secondoftwo}%
\providecommand \translation [1]{[#1]}%
\providecommand \BibitemOpen [0]{}%
\providecommand \bibitemStop [0]{}%
\providecommand \bibitemNoStop [0]{.\EOS\space}%
\providecommand \EOS [0]{\spacefactor3000\relax}%
\providecommand \BibitemShut  [1]{\csname bibitem#1\endcsname}%
\let\auto@bib@innerbib\@empty
\bibitem [{\citenamefont {Hualei}\ \emph {et~al.}(2023)\citenamefont {Hualei}, \citenamefont {Huo}, \citenamefont {Hu}, \citenamefont {Li}, \citenamefont {Liu}, \citenamefont {Han}, \citenamefont {Tang}, \citenamefont {Mao}, \citenamefont {Yang}, \citenamefont {Wang}, \citenamefont {Cheng}, \citenamefont {Yao}, \citenamefont {Zhang},\ and\ \citenamefont {Wang}}]{Nature621}%
  \BibitemOpen
  \bibfield  {author} {\bibinfo {author} {\bibfnamefont {S.}~\bibnamefont {Hualei}}, \bibinfo {author} {\bibfnamefont {M.}~\bibnamefont {Huo}}, \bibinfo {author} {\bibfnamefont {X.}~\bibnamefont {Hu}}, \bibinfo {author} {\bibfnamefont {J.}~\bibnamefont {Li}}, \bibinfo {author} {\bibfnamefont {Z.}~\bibnamefont {Liu}}, \bibinfo {author} {\bibfnamefont {Y.}~\bibnamefont {Han}}, \bibinfo {author} {\bibfnamefont {L.}~\bibnamefont {Tang}}, \bibinfo {author} {\bibfnamefont {Z.}~\bibnamefont {Mao}}, \bibinfo {author} {\bibfnamefont {P.}~\bibnamefont {Yang}}, \bibinfo {author} {\bibfnamefont {B.}~\bibnamefont {Wang}}, \bibinfo {author} {\bibfnamefont {J.}~\bibnamefont {Cheng}}, \bibinfo {author} {\bibfnamefont {D.-X.}\ \bibnamefont {Yao}}, \bibinfo {author} {\bibfnamefont {G.-M.}\ \bibnamefont {Zhang}},\ and\ \bibinfo {author} {\bibfnamefont {M.}~\bibnamefont {Wang}},\ }\href {https://doi.org/10.1038/s41586-023-06408-7} {\bibfield  {journal} {\bibinfo  {journal} {Nature}\ }\textbf {\bibinfo {volume} {621}},\ \bibinfo
  {pages} {493} (\bibinfo {year} {2023})}\BibitemShut {NoStop}%
\bibitem [{\citenamefont {Zhang}\ \emph {et~al.}(2024{\natexlab{a}})\citenamefont {Zhang}, \citenamefont {Su}, \citenamefont {Huang}, \citenamefont {Shan}, \citenamefont {Sun}, \citenamefont {Huo}, \citenamefont {Ye}, \citenamefont {Zhang}, \citenamefont {Yang}, \citenamefont {Xu}, \citenamefont {Su}, \citenamefont {Li}, \citenamefont {Smidman}, \citenamefont {Wang}, \citenamefont {Jiao},\ and\ \citenamefont {Yuan}}]{Zhang2024}%
  \BibitemOpen
  \bibfield  {author} {\bibinfo {author} {\bibfnamefont {Y.}~\bibnamefont {Zhang}}, \bibinfo {author} {\bibfnamefont {D.}~\bibnamefont {Su}}, \bibinfo {author} {\bibfnamefont {Y.}~\bibnamefont {Huang}}, \bibinfo {author} {\bibfnamefont {Z.}~\bibnamefont {Shan}}, \bibinfo {author} {\bibfnamefont {H.}~\bibnamefont {Sun}}, \bibinfo {author} {\bibfnamefont {M.}~\bibnamefont {Huo}}, \bibinfo {author} {\bibfnamefont {K.}~\bibnamefont {Ye}}, \bibinfo {author} {\bibfnamefont {J.}~\bibnamefont {Zhang}}, \bibinfo {author} {\bibfnamefont {Z.}~\bibnamefont {Yang}}, \bibinfo {author} {\bibfnamefont {Y.}~\bibnamefont {Xu}}, \bibinfo {author} {\bibfnamefont {Y.}~\bibnamefont {Su}}, \bibinfo {author} {\bibfnamefont {R.}~\bibnamefont {Li}}, \bibinfo {author} {\bibfnamefont {M.}~\bibnamefont {Smidman}}, \bibinfo {author} {\bibfnamefont {M.}~\bibnamefont {Wang}}, \bibinfo {author} {\bibfnamefont {L.}~\bibnamefont {Jiao}},\ and\ \bibinfo {author} {\bibfnamefont {H.}~\bibnamefont {Yuan}},\ }\href
  {https://doi.org/10.1038/s41567-024-02515-y} {\bibfield  {journal} {\bibinfo  {journal} {Nature Physics}\ }\textbf {\bibinfo {volume} {20}},\ \bibinfo {pages} {1269} (\bibinfo {year} {2024}{\natexlab{a}})}\BibitemShut {NoStop}%
\bibitem [{\citenamefont {Puphal}\ \emph {et~al.}(2024)\citenamefont {Puphal}, \citenamefont {Reiss}, \citenamefont {Enderlein}, \citenamefont {Wu}, \citenamefont {Khaliullin}, \citenamefont {Sundaramurthy}, \citenamefont {Priessnitz}, \citenamefont {Knauft}, \citenamefont {Suthar}, \citenamefont {Richter}, \citenamefont {Isobe}, \citenamefont {van Aken}, \citenamefont {Takagi}, \citenamefont {Keimer}, \citenamefont {Suyolcu}, \citenamefont {Wehinger}, \citenamefont {Hansmann},\ and\ \citenamefont {Hepting}}]{PhysRevLett.133.146002}%
  \BibitemOpen
  \bibfield  {author} {\bibinfo {author} {\bibfnamefont {P.}~\bibnamefont {Puphal}}, \bibinfo {author} {\bibfnamefont {P.}~\bibnamefont {Reiss}}, \bibinfo {author} {\bibfnamefont {N.}~\bibnamefont {Enderlein}}, \bibinfo {author} {\bibfnamefont {Y.-M.}\ \bibnamefont {Wu}}, \bibinfo {author} {\bibfnamefont {G.}~\bibnamefont {Khaliullin}}, \bibinfo {author} {\bibfnamefont {V.}~\bibnamefont {Sundaramurthy}}, \bibinfo {author} {\bibfnamefont {T.}~\bibnamefont {Priessnitz}}, \bibinfo {author} {\bibfnamefont {M.}~\bibnamefont {Knauft}}, \bibinfo {author} {\bibfnamefont {A.}~\bibnamefont {Suthar}}, \bibinfo {author} {\bibfnamefont {L.}~\bibnamefont {Richter}}, \bibinfo {author} {\bibfnamefont {M.}~\bibnamefont {Isobe}}, \bibinfo {author} {\bibfnamefont {P.~A.}\ \bibnamefont {van Aken}}, \bibinfo {author} {\bibfnamefont {H.}~\bibnamefont {Takagi}}, \bibinfo {author} {\bibfnamefont {B.}~\bibnamefont {Keimer}}, \bibinfo {author} {\bibfnamefont {Y.~E.}\ \bibnamefont {Suyolcu}}, \bibinfo {author} {\bibfnamefont
  {B.}~\bibnamefont {Wehinger}}, \bibinfo {author} {\bibfnamefont {P.}~\bibnamefont {Hansmann}},\ and\ \bibinfo {author} {\bibfnamefont {M.}~\bibnamefont {Hepting}},\ }\href {https://doi.org/10.1103/PhysRevLett.133.146002} {\bibfield  {journal} {\bibinfo  {journal} {Phys. Rev. Lett.}\ }\textbf {\bibinfo {volume} {133}},\ \bibinfo {pages} {146002} (\bibinfo {year} {2024})}\BibitemShut {NoStop}%
\bibitem [{\citenamefont {Wang}\ \emph {et~al.}(2024{\natexlab{a}})\citenamefont {Wang}, \citenamefont {Wang}, \citenamefont {Shen}, \citenamefont {Hou}, \citenamefont {Ma}, \citenamefont {Shi}, \citenamefont {Ren}, \citenamefont {Gu}, \citenamefont {Ma}, \citenamefont {Yang}, \citenamefont {Liu}, \citenamefont {Guo}, \citenamefont {Sun}, \citenamefont {Zhang}, \citenamefont {Calder}, \citenamefont {Yan}, \citenamefont {Wang}, \citenamefont {Uwatoko},\ and\ \citenamefont {Cheng}}]{PhysRevX.14.011040}%
  \BibitemOpen
  \bibfield  {author} {\bibinfo {author} {\bibfnamefont {G.}~\bibnamefont {Wang}}, \bibinfo {author} {\bibfnamefont {N.~N.}\ \bibnamefont {Wang}}, \bibinfo {author} {\bibfnamefont {X.~L.}\ \bibnamefont {Shen}}, \bibinfo {author} {\bibfnamefont {J.}~\bibnamefont {Hou}}, \bibinfo {author} {\bibfnamefont {L.}~\bibnamefont {Ma}}, \bibinfo {author} {\bibfnamefont {L.~F.}\ \bibnamefont {Shi}}, \bibinfo {author} {\bibfnamefont {Z.~A.}\ \bibnamefont {Ren}}, \bibinfo {author} {\bibfnamefont {Y.~D.}\ \bibnamefont {Gu}}, \bibinfo {author} {\bibfnamefont {H.~M.}\ \bibnamefont {Ma}}, \bibinfo {author} {\bibfnamefont {P.~T.}\ \bibnamefont {Yang}}, \bibinfo {author} {\bibfnamefont {Z.~Y.}\ \bibnamefont {Liu}}, \bibinfo {author} {\bibfnamefont {H.~Z.}\ \bibnamefont {Guo}}, \bibinfo {author} {\bibfnamefont {J.~P.}\ \bibnamefont {Sun}}, \bibinfo {author} {\bibfnamefont {G.~M.}\ \bibnamefont {Zhang}}, \bibinfo {author} {\bibfnamefont {S.}~\bibnamefont {Calder}}, \bibinfo {author} {\bibfnamefont {J.-Q.}\ \bibnamefont {Yan}},
  \bibinfo {author} {\bibfnamefont {B.~S.}\ \bibnamefont {Wang}}, \bibinfo {author} {\bibfnamefont {Y.}~\bibnamefont {Uwatoko}},\ and\ \bibinfo {author} {\bibfnamefont {J.-G.}\ \bibnamefont {Cheng}},\ }\href {https://doi.org/10.1103/PhysRevX.14.011040} {\bibfield  {journal} {\bibinfo  {journal} {Phys. Rev. X}\ }\textbf {\bibinfo {volume} {14}},\ \bibinfo {pages} {011040} (\bibinfo {year} {2024}{\natexlab{a}})}\BibitemShut {NoStop}%
\bibitem [{\citenamefont {Zhu}\ \emph {et~al.}(2024)\citenamefont {Zhu}, \citenamefont {Peng}, \citenamefont {Zhang}, \citenamefont {Pan}, \citenamefont {Chen}, \citenamefont {Chen}, \citenamefont {Ren}, \citenamefont {Liu}, \citenamefont {Hao}, \citenamefont {Li}, \citenamefont {Xing}, \citenamefont {Lan}, \citenamefont {Han}, \citenamefont {Wang}, \citenamefont {Jia}, \citenamefont {Wo}, \citenamefont {Gu}, \citenamefont {Gu}, \citenamefont {Ji}, \citenamefont {Wang}, \citenamefont {Gou}, \citenamefont {Shen}, \citenamefont {Ying}, \citenamefont {Chen}, \citenamefont {Yang}, \citenamefont {Cao}, \citenamefont {Zheng}, \citenamefont {Zeng}, \citenamefont {gang Guo},\ and\ \citenamefont {Zhao}}]{Nature631}%
  \BibitemOpen
  \bibfield  {author} {\bibinfo {author} {\bibfnamefont {Y.}~\bibnamefont {Zhu}}, \bibinfo {author} {\bibfnamefont {D.}~\bibnamefont {Peng}}, \bibinfo {author} {\bibfnamefont {E.}~\bibnamefont {Zhang}}, \bibinfo {author} {\bibfnamefont {B.}~\bibnamefont {Pan}}, \bibinfo {author} {\bibfnamefont {X.}~\bibnamefont {Chen}}, \bibinfo {author} {\bibfnamefont {L.}~\bibnamefont {Chen}}, \bibinfo {author} {\bibfnamefont {H.}~\bibnamefont {Ren}}, \bibinfo {author} {\bibfnamefont {F.}~\bibnamefont {Liu}}, \bibinfo {author} {\bibfnamefont {Y.}~\bibnamefont {Hao}}, \bibinfo {author} {\bibfnamefont {N.}~\bibnamefont {Li}}, \bibinfo {author} {\bibfnamefont {Z.}~\bibnamefont {Xing}}, \bibinfo {author} {\bibfnamefont {F.}~\bibnamefont {Lan}}, \bibinfo {author} {\bibfnamefont {J.}~\bibnamefont {Han}}, \bibinfo {author} {\bibfnamefont {J.}~\bibnamefont {Wang}}, \bibinfo {author} {\bibfnamefont {D.}~\bibnamefont {Jia}}, \bibinfo {author} {\bibfnamefont {H.}~\bibnamefont {Wo}}, \bibinfo {author} {\bibfnamefont {Y.}~\bibnamefont
  {Gu}}, \bibinfo {author} {\bibfnamefont {Y.}~\bibnamefont {Gu}}, \bibinfo {author} {\bibfnamefont {L.}~\bibnamefont {Ji}}, \bibinfo {author} {\bibfnamefont {W.}~\bibnamefont {Wang}}, \bibinfo {author} {\bibfnamefont {H.}~\bibnamefont {Gou}}, \bibinfo {author} {\bibfnamefont {Y.}~\bibnamefont {Shen}}, \bibinfo {author} {\bibfnamefont {T.}~\bibnamefont {Ying}}, \bibinfo {author} {\bibfnamefont {X.}~\bibnamefont {Chen}}, \bibinfo {author} {\bibfnamefont {W.}~\bibnamefont {Yang}}, \bibinfo {author} {\bibfnamefont {H.}~\bibnamefont {Cao}}, \bibinfo {author} {\bibfnamefont {C.}~\bibnamefont {Zheng}}, \bibinfo {author} {\bibfnamefont {Q.}~\bibnamefont {Zeng}}, \bibinfo {author} {\bibfnamefont {J.}~\bibnamefont {gang Guo}},\ and\ \bibinfo {author} {\bibfnamefont {J.}~\bibnamefont {Zhao}},\ }\href {https://doi.org/10.1038/s41586-024-07553-3} {\bibfield  {journal} {\bibinfo  {journal} {Nature}\ }\textbf {\bibinfo {volume} {631}},\ \bibinfo {pages} {531} (\bibinfo {year} {2024})}\BibitemShut {NoStop}%
\bibitem [{\citenamefont {Kakoi}\ \emph {et~al.}(2024)\citenamefont {Kakoi}, \citenamefont {Oi}, \citenamefont {Ohshita}, \citenamefont {Yashima}, \citenamefont {Kuroki}, \citenamefont {Kato}, \citenamefont {Takahashi}, \citenamefont {Ishiwata}, \citenamefont {Adachi}, \citenamefont {Hatada}, \citenamefont {Uda},\ and\ \citenamefont {Mukuda}}]{doi:10.7566/JPSJ.93.053702}%
  \BibitemOpen
  \bibfield  {author} {\bibinfo {author} {\bibfnamefont {M.}~\bibnamefont {Kakoi}}, \bibinfo {author} {\bibfnamefont {T.}~\bibnamefont {Oi}}, \bibinfo {author} {\bibfnamefont {Y.}~\bibnamefont {Ohshita}}, \bibinfo {author} {\bibfnamefont {M.}~\bibnamefont {Yashima}}, \bibinfo {author} {\bibfnamefont {K.}~\bibnamefont {Kuroki}}, \bibinfo {author} {\bibfnamefont {T.}~\bibnamefont {Kato}}, \bibinfo {author} {\bibfnamefont {H.}~\bibnamefont {Takahashi}}, \bibinfo {author} {\bibfnamefont {S.}~\bibnamefont {Ishiwata}}, \bibinfo {author} {\bibfnamefont {Y.}~\bibnamefont {Adachi}}, \bibinfo {author} {\bibfnamefont {N.}~\bibnamefont {Hatada}}, \bibinfo {author} {\bibfnamefont {T.}~\bibnamefont {Uda}},\ and\ \bibinfo {author} {\bibfnamefont {H.}~\bibnamefont {Mukuda}},\ }\href {https://doi.org/10.7566/JPSJ.93.053702} {\bibfield  {journal} {\bibinfo  {journal} {Journal of the Physical Society of Japan}\ }\textbf {\bibinfo {volume} {93}},\ \bibinfo {pages} {053702} (\bibinfo {year} {2024})}\BibitemShut {NoStop}%
\bibitem [{\citenamefont {Sakakibara}\ \emph {et~al.}(2024{\natexlab{a}})\citenamefont {Sakakibara}, \citenamefont {Ochi}, \citenamefont {Nagata}, \citenamefont {Ueki}, \citenamefont {Sakurai}, \citenamefont {Matsumoto}, \citenamefont {Terashima}, \citenamefont {Hirose}, \citenamefont {Ohta}, \citenamefont {Kato}, \citenamefont {Takano},\ and\ \citenamefont {Kuroki}}]{PhysRevB.109.144511}%
  \BibitemOpen
  \bibfield  {author} {\bibinfo {author} {\bibfnamefont {H.}~\bibnamefont {Sakakibara}}, \bibinfo {author} {\bibfnamefont {M.}~\bibnamefont {Ochi}}, \bibinfo {author} {\bibfnamefont {H.}~\bibnamefont {Nagata}}, \bibinfo {author} {\bibfnamefont {Y.}~\bibnamefont {Ueki}}, \bibinfo {author} {\bibfnamefont {H.}~\bibnamefont {Sakurai}}, \bibinfo {author} {\bibfnamefont {R.}~\bibnamefont {Matsumoto}}, \bibinfo {author} {\bibfnamefont {K.}~\bibnamefont {Terashima}}, \bibinfo {author} {\bibfnamefont {K.}~\bibnamefont {Hirose}}, \bibinfo {author} {\bibfnamefont {H.}~\bibnamefont {Ohta}}, \bibinfo {author} {\bibfnamefont {M.}~\bibnamefont {Kato}}, \bibinfo {author} {\bibfnamefont {Y.}~\bibnamefont {Takano}},\ and\ \bibinfo {author} {\bibfnamefont {K.}~\bibnamefont {Kuroki}},\ }\href {https://doi.org/10.1103/PhysRevB.109.144511} {\bibfield  {journal} {\bibinfo  {journal} {Phys. Rev. B}\ }\textbf {\bibinfo {volume} {109}},\ \bibinfo {pages} {144511} (\bibinfo {year} {2024}{\natexlab{a}})}\BibitemShut {NoStop}%
\bibitem [{\citenamefont {Li}\ \emph {et~al.}(2024)\citenamefont {Li}, \citenamefont {Zhang}, \citenamefont {Xiang}, \citenamefont {Zhang}, \citenamefont {Zhu},\ and\ \citenamefont {Wen}}]{QingLi:17401}%
  \BibitemOpen
  \bibfield  {author} {\bibinfo {author} {\bibfnamefont {Q.}~\bibnamefont {Li}}, \bibinfo {author} {\bibfnamefont {Y.-J.}\ \bibnamefont {Zhang}}, \bibinfo {author} {\bibfnamefont {Z.-N.}\ \bibnamefont {Xiang}}, \bibinfo {author} {\bibfnamefont {Y.}~\bibnamefont {Zhang}}, \bibinfo {author} {\bibfnamefont {X.}~\bibnamefont {Zhu}},\ and\ \bibinfo {author} {\bibfnamefont {H.-H.}\ \bibnamefont {Wen}},\ }\href {https://doi.org/10.1088/0256-307X/41/1/017401} {\bibfield  {journal} {\bibinfo  {journal} {Chinese Physics Letters}\ }\textbf {\bibinfo {volume} {41}},\ \bibinfo {eid} {017401} (\bibinfo {year} {2024})}\BibitemShut {NoStop}%
\bibitem [{\citenamefont {Zhang}\ \emph {et~al.}(2024{\natexlab{b}})\citenamefont {Zhang}, \citenamefont {Pei}, \citenamefont {Du}, \citenamefont {Hu}, \citenamefont {Cao}, \citenamefont {Wang}, \citenamefont {Wu}, \citenamefont {Li}, \citenamefont {Liu}, \citenamefont {Wen}, \citenamefont {Zhao}, \citenamefont {Li}, \citenamefont {Cao}, \citenamefont {Zhu}, \citenamefont {Zhang}, \citenamefont {Yu}, \citenamefont {Cheng}, \citenamefont {Zhang}, \citenamefont {Li}, \citenamefont {Zhao}, \citenamefont {Chen}, \citenamefont {Guo}, \citenamefont {Wu}, \citenamefont {Yang}, \citenamefont {Yan}, \citenamefont {Yang},\ and\ \citenamefont {Qi}}]{zhang2024superconductivitytrilayernickelatela4ni3o10}%
  \BibitemOpen
  \bibfield  {author} {\bibinfo {author} {\bibfnamefont {M.}~\bibnamefont {Zhang}}, \bibinfo {author} {\bibfnamefont {C.}~\bibnamefont {Pei}}, \bibinfo {author} {\bibfnamefont {X.}~\bibnamefont {Du}}, \bibinfo {author} {\bibfnamefont {W.}~\bibnamefont {Hu}}, \bibinfo {author} {\bibfnamefont {Y.}~\bibnamefont {Cao}}, \bibinfo {author} {\bibfnamefont {Q.}~\bibnamefont {Wang}}, \bibinfo {author} {\bibfnamefont {J.}~\bibnamefont {Wu}}, \bibinfo {author} {\bibfnamefont {Y.}~\bibnamefont {Li}}, \bibinfo {author} {\bibfnamefont {H.}~\bibnamefont {Liu}}, \bibinfo {author} {\bibfnamefont {C.}~\bibnamefont {Wen}}, \bibinfo {author} {\bibfnamefont {Y.}~\bibnamefont {Zhao}}, \bibinfo {author} {\bibfnamefont {C.}~\bibnamefont {Li}}, \bibinfo {author} {\bibfnamefont {W.}~\bibnamefont {Cao}}, \bibinfo {author} {\bibfnamefont {S.}~\bibnamefont {Zhu}}, \bibinfo {author} {\bibfnamefont {Q.}~\bibnamefont {Zhang}}, \bibinfo {author} {\bibfnamefont {N.}~\bibnamefont {Yu}}, \bibinfo {author} {\bibfnamefont {P.}~\bibnamefont
  {Cheng}}, \bibinfo {author} {\bibfnamefont {L.}~\bibnamefont {Zhang}}, \bibinfo {author} {\bibfnamefont {Z.}~\bibnamefont {Li}}, \bibinfo {author} {\bibfnamefont {J.}~\bibnamefont {Zhao}}, \bibinfo {author} {\bibfnamefont {Y.}~\bibnamefont {Chen}}, \bibinfo {author} {\bibfnamefont {H.}~\bibnamefont {Guo}}, \bibinfo {author} {\bibfnamefont {C.}~\bibnamefont {Wu}}, \bibinfo {author} {\bibfnamefont {F.}~\bibnamefont {Yang}}, \bibinfo {author} {\bibfnamefont {S.}~\bibnamefont {Yan}}, \bibinfo {author} {\bibfnamefont {L.}~\bibnamefont {Yang}},\ and\ \bibinfo {author} {\bibfnamefont {Y.}~\bibnamefont {Qi}},\ }\href {https://arxiv.org/abs/2311.07423} {\bibinfo {title} {Superconductivity in trilayer nickelate $\rm{La}_{4}{Ni}_{3}{O}_{10}$ under pressure}} (\bibinfo {year} {2024}{\natexlab{b}}),\ \Eprint {https://arxiv.org/abs/2311.07423} {arXiv:2311.07423 [cond-mat.supr-con]} \BibitemShut {NoStop}%
\bibitem [{\citenamefont {{Seo}}\ \emph {et~al.}(1996)\citenamefont {{Seo}}, \citenamefont {{Liang}}, \citenamefont {{Whangbo}}, \citenamefont {{Zhang}},\ and\ \citenamefont {{Greenblatt}}}]{doi:10.1021/ic960379j}%
  \BibitemOpen
  \bibfield  {author} {\bibinfo {author} {\bibfnamefont {D.-K.}\ \bibnamefont {{Seo}}}, \bibinfo {author} {\bibfnamefont {W.}~\bibnamefont {{Liang}}}, \bibinfo {author} {\bibfnamefont {M.-H.}\ \bibnamefont {{Whangbo}}}, \bibinfo {author} {\bibfnamefont {Z.}~\bibnamefont {{Zhang}}},\ and\ \bibinfo {author} {\bibfnamefont {M.}~\bibnamefont {{Greenblatt}}},\ }\href {https://doi.org/10.1021/ic960379j} {\bibfield  {journal} {\bibinfo  {journal} {Inorganic Chemistry}\ }\textbf {\bibinfo {volume} {35}},\ \bibinfo {pages} {6396} (\bibinfo {year} {1996})},\ \bibinfo {note} {pMID: 11666785}\BibitemShut {NoStop}%
\bibitem [{\citenamefont {Ni}\ \emph {et~al.}(2025)\citenamefont {Ni}, \citenamefont {Ji}, \citenamefont {He}, \citenamefont {Xie}, \citenamefont {Yao}, \citenamefont {Wang},\ and\ \citenamefont {Cao}}]{Ni2025SpinDensityWave}%
  \BibitemOpen
  \bibfield  {author} {\bibinfo {author} {\bibfnamefont {X.-S.}\ \bibnamefont {Ni}}, \bibinfo {author} {\bibfnamefont {Y.}~\bibnamefont {Ji}}, \bibinfo {author} {\bibfnamefont {L.}~\bibnamefont {He}}, \bibinfo {author} {\bibfnamefont {T.}~\bibnamefont {Xie}}, \bibinfo {author} {\bibfnamefont {D.-X.}\ \bibnamefont {Yao}}, \bibinfo {author} {\bibfnamefont {M.}~\bibnamefont {Wang}},\ and\ \bibinfo {author} {\bibfnamefont {K.}~\bibnamefont {Cao}},\ }\href {https://doi.org/10.1038/s41535-025-00740-z} {\bibfield  {journal} {\bibinfo  {journal} {npj Quantum Mater.}\ }\textbf {\bibinfo {volume} {10}},\ \bibinfo {pages} {17} (\bibinfo {year} {2025})}\BibitemShut {NoStop}%
\bibitem [{\citenamefont {Luo}\ \emph {et~al.}(2023)\citenamefont {Luo}, \citenamefont {Hu}, \citenamefont {Wang}, \citenamefont {W\'u},\ and\ \citenamefont {Yao}}]{PhysRevLett.131.126001}%
  \BibitemOpen
  \bibfield  {author} {\bibinfo {author} {\bibfnamefont {Z.}~\bibnamefont {Luo}}, \bibinfo {author} {\bibfnamefont {X.}~\bibnamefont {Hu}}, \bibinfo {author} {\bibfnamefont {M.}~\bibnamefont {Wang}}, \bibinfo {author} {\bibfnamefont {W.}~\bibnamefont {W\'u}},\ and\ \bibinfo {author} {\bibfnamefont {D.-X.}\ \bibnamefont {Yao}},\ }\href {https://doi.org/10.1103/PhysRevLett.131.126001} {\bibfield  {journal} {\bibinfo  {journal} {Phys. Rev. Lett.}\ }\textbf {\bibinfo {volume} {131}},\ \bibinfo {pages} {126001} (\bibinfo {year} {2023})}\BibitemShut {NoStop}%
\bibitem [{\citenamefont {Christiansson}\ \emph {et~al.}(2023)\citenamefont {Christiansson}, \citenamefont {Petocchi},\ and\ \citenamefont {Werner}}]{PhysRevLett.131.206501}%
  \BibitemOpen
  \bibfield  {author} {\bibinfo {author} {\bibfnamefont {V.}~\bibnamefont {Christiansson}}, \bibinfo {author} {\bibfnamefont {F.}~\bibnamefont {Petocchi}},\ and\ \bibinfo {author} {\bibfnamefont {P.}~\bibnamefont {Werner}},\ }\href {https://doi.org/10.1103/PhysRevLett.131.206501} {\bibfield  {journal} {\bibinfo  {journal} {Phys. Rev. Lett.}\ }\textbf {\bibinfo {volume} {131}},\ \bibinfo {pages} {206501} (\bibinfo {year} {2023})}\BibitemShut {NoStop}%
\bibitem [{\citenamefont {Chen}\ \emph {et~al.}(2024{\natexlab{a}})\citenamefont {Chen}, \citenamefont {Choi}, \citenamefont {Jiang}, \citenamefont {Mei}, \citenamefont {Jiang}, \citenamefont {Li}, \citenamefont {Agrestini}, \citenamefont {Garcia-Fernandez}, \citenamefont {Huang}, \citenamefont {Sun}, \citenamefont {Shen}, \citenamefont {Wang}, \citenamefont {Hu}, \citenamefont {Lu}, \citenamefont {Zhou},\ and\ \citenamefont {Feng}}]{Chen2024}%
  \BibitemOpen
  \bibfield  {author} {\bibinfo {author} {\bibfnamefont {X.}~\bibnamefont {Chen}}, \bibinfo {author} {\bibfnamefont {J.}~\bibnamefont {Choi}}, \bibinfo {author} {\bibfnamefont {Z.}~\bibnamefont {Jiang}}, \bibinfo {author} {\bibfnamefont {J.}~\bibnamefont {Mei}}, \bibinfo {author} {\bibfnamefont {K.}~\bibnamefont {Jiang}}, \bibinfo {author} {\bibfnamefont {J.}~\bibnamefont {Li}}, \bibinfo {author} {\bibfnamefont {S.}~\bibnamefont {Agrestini}}, \bibinfo {author} {\bibfnamefont {M.}~\bibnamefont {Garcia-Fernandez}}, \bibinfo {author} {\bibfnamefont {X.}~\bibnamefont {Huang}}, \bibinfo {author} {\bibfnamefont {H.}~\bibnamefont {Sun}}, \bibinfo {author} {\bibfnamefont {D.}~\bibnamefont {Shen}}, \bibinfo {author} {\bibfnamefont {M.}~\bibnamefont {Wang}}, \bibinfo {author} {\bibfnamefont {J.}~\bibnamefont {Hu}}, \bibinfo {author} {\bibfnamefont {Y.}~\bibnamefont {Lu}}, \bibinfo {author} {\bibfnamefont {K.-J.}\ \bibnamefont {Zhou}},\ and\ \bibinfo {author} {\bibfnamefont {D.}~\bibnamefont {Feng}},\ }\href
  {https://doi.org/10.1038/s41467-024-53863-5} {\bibfield  {journal} {\bibinfo  {journal} {Nature Communications}\ }\textbf {\bibinfo {volume} {15}},\ \bibinfo {pages} {9597} (\bibinfo {year} {2024}{\natexlab{a}})}\BibitemShut {NoStop}%
\bibitem [{\citenamefont {Xie}\ \emph {et~al.}(2024)\citenamefont {Xie}, \citenamefont {Huo}, \citenamefont {Ni}, \citenamefont {Shen}, \citenamefont {Huang}, \citenamefont {Sun}, \citenamefont {Walker}, \citenamefont {Adroja}, \citenamefont {Yu}, \citenamefont {Shen}, \citenamefont {He}, \citenamefont {Cao},\ and\ \citenamefont {Wang}}]{xie20243221}%
  \BibitemOpen
  \bibfield  {author} {\bibinfo {author} {\bibfnamefont {T.}~\bibnamefont {Xie}}, \bibinfo {author} {\bibfnamefont {M.}~\bibnamefont {Huo}}, \bibinfo {author} {\bibfnamefont {X.}~\bibnamefont {Ni}}, \bibinfo {author} {\bibfnamefont {F.}~\bibnamefont {Shen}}, \bibinfo {author} {\bibfnamefont {X.}~\bibnamefont {Huang}}, \bibinfo {author} {\bibfnamefont {H.}~\bibnamefont {Sun}}, \bibinfo {author} {\bibfnamefont {H.~C.}\ \bibnamefont {Walker}}, \bibinfo {author} {\bibfnamefont {D.}~\bibnamefont {Adroja}}, \bibinfo {author} {\bibfnamefont {D.}~\bibnamefont {Yu}}, \bibinfo {author} {\bibfnamefont {B.}~\bibnamefont {Shen}}, \bibinfo {author} {\bibfnamefont {L.}~\bibnamefont {He}}, \bibinfo {author} {\bibfnamefont {K.}~\bibnamefont {Cao}},\ and\ \bibinfo {author} {\bibfnamefont {M.}~\bibnamefont {Wang}},\ }\href {https://doi.org/https://doi.org/10.1016/j.scib.2024.07.030} {\bibfield  {journal} {\bibinfo  {journal} {Science Bulletin}\ }\textbf {\bibinfo {volume} {69}},\ \bibinfo {pages} {3221} (\bibinfo {year}
  {2024})}\BibitemShut {NoStop}%
\bibitem [{\citenamefont {Zhao}\ \emph {et~al.}(2025)\citenamefont {Zhao}, \citenamefont {Zhou}, \citenamefont {Huo}, \citenamefont {Wang}, \citenamefont {Nie}, \citenamefont {Yang}, \citenamefont {Ying}, \citenamefont {Wang}, \citenamefont {Wu},\ and\ \citenamefont {Chen}}]{ZHAO2025}%
  \BibitemOpen
  \bibfield  {author} {\bibinfo {author} {\bibfnamefont {D.}~\bibnamefont {Zhao}}, \bibinfo {author} {\bibfnamefont {Y.}~\bibnamefont {Zhou}}, \bibinfo {author} {\bibfnamefont {M.}~\bibnamefont {Huo}}, \bibinfo {author} {\bibfnamefont {Y.}~\bibnamefont {Wang}}, \bibinfo {author} {\bibfnamefont {L.}~\bibnamefont {Nie}}, \bibinfo {author} {\bibfnamefont {Y.}~\bibnamefont {Yang}}, \bibinfo {author} {\bibfnamefont {J.}~\bibnamefont {Ying}}, \bibinfo {author} {\bibfnamefont {M.}~\bibnamefont {Wang}}, \bibinfo {author} {\bibfnamefont {T.}~\bibnamefont {Wu}},\ and\ \bibinfo {author} {\bibfnamefont {X.}~\bibnamefont {Chen}},\ }\href {https://doi.org/https://doi.org/10.1016/j.scib.2025.02.019} {\bibfield  {journal} {\bibinfo  {journal} {Science Bulletin}\ }\textbf {\bibinfo {volume} {70}},\ \bibinfo {pages} {1239} (\bibinfo {year} {2025})}\BibitemShut {NoStop}%
\bibitem [{\citenamefont {Gupta}\ \emph {et~al.}(2024)\citenamefont {Gupta}, \citenamefont {Gong}, \citenamefont {Wu}, \citenamefont {Kang}, \citenamefont {Parzyck}, \citenamefont {Gregory}, \citenamefont {Costa}, \citenamefont {Sutarto}, \citenamefont {Sarker}, \citenamefont {Singer}, \citenamefont {Schlom}, \citenamefont {Shen},\ and\ \citenamefont {Hawthorn}}]{gupta2024anisotropicspinstripedomains}%
  \BibitemOpen
  \bibfield  {author} {\bibinfo {author} {\bibfnamefont {N.~K.}\ \bibnamefont {Gupta}}, \bibinfo {author} {\bibfnamefont {R.}~\bibnamefont {Gong}}, \bibinfo {author} {\bibfnamefont {Y.}~\bibnamefont {Wu}}, \bibinfo {author} {\bibfnamefont {M.}~\bibnamefont {Kang}}, \bibinfo {author} {\bibfnamefont {C.~T.}\ \bibnamefont {Parzyck}}, \bibinfo {author} {\bibfnamefont {B.~Z.}\ \bibnamefont {Gregory}}, \bibinfo {author} {\bibfnamefont {N.}~\bibnamefont {Costa}}, \bibinfo {author} {\bibfnamefont {R.}~\bibnamefont {Sutarto}}, \bibinfo {author} {\bibfnamefont {S.}~\bibnamefont {Sarker}}, \bibinfo {author} {\bibfnamefont {A.}~\bibnamefont {Singer}}, \bibinfo {author} {\bibfnamefont {D.~G.}\ \bibnamefont {Schlom}}, \bibinfo {author} {\bibfnamefont {K.~M.}\ \bibnamefont {Shen}},\ and\ \bibinfo {author} {\bibfnamefont {D.~G.}\ \bibnamefont {Hawthorn}},\ }\href {https://arxiv.org/abs/2409.03210} {\bibinfo {title} {Anisotropic spin stripe domains in bilayer $\rm{La}_{3}{Ni}_{2}{O}_{7}$}} (\bibinfo {year} {2024}),\ \Eprint
  {https://arxiv.org/abs/2409.03210} {arXiv:2409.03210 [cond-mat.supr-con]} \BibitemShut {NoStop}%
\bibitem [{\citenamefont {Chen}\ \emph {et~al.}(2024{\natexlab{b}})\citenamefont {Chen}, \citenamefont {Liu}, \citenamefont {Jiao}, \citenamefont {Zou}, \citenamefont {Jiang}, \citenamefont {Li}, \citenamefont {Luo}, \citenamefont {Wu}, \citenamefont {Zhang}, \citenamefont {Guo},\ and\ \citenamefont {Shu}}]{PhysRevLett.132.256503}%
  \BibitemOpen
  \bibfield  {author} {\bibinfo {author} {\bibfnamefont {K.}~\bibnamefont {Chen}}, \bibinfo {author} {\bibfnamefont {X.}~\bibnamefont {Liu}}, \bibinfo {author} {\bibfnamefont {J.}~\bibnamefont {Jiao}}, \bibinfo {author} {\bibfnamefont {M.}~\bibnamefont {Zou}}, \bibinfo {author} {\bibfnamefont {C.}~\bibnamefont {Jiang}}, \bibinfo {author} {\bibfnamefont {X.}~\bibnamefont {Li}}, \bibinfo {author} {\bibfnamefont {Y.}~\bibnamefont {Luo}}, \bibinfo {author} {\bibfnamefont {Q.}~\bibnamefont {Wu}}, \bibinfo {author} {\bibfnamefont {N.}~\bibnamefont {Zhang}}, \bibinfo {author} {\bibfnamefont {Y.}~\bibnamefont {Guo}},\ and\ \bibinfo {author} {\bibfnamefont {L.}~\bibnamefont {Shu}},\ }\href {https://doi.org/10.1103/PhysRevLett.132.256503} {\bibfield  {journal} {\bibinfo  {journal} {Phys. Rev. Lett.}\ }\textbf {\bibinfo {volume} {132}},\ \bibinfo {pages} {256503} (\bibinfo {year} {2024}{\natexlab{b}})}\BibitemShut {NoStop}%
\bibitem [{\citenamefont {Meng}\ \emph {et~al.}(2024)\citenamefont {Meng}, \citenamefont {Yang}, \citenamefont {Sun}, \citenamefont {Zhang}, \citenamefont {Luo}, \citenamefont {Wang}, \citenamefont {Hong}, \citenamefont {Wang},\ and\ \citenamefont {Yu}}]{meng2024densitywavelikegapevolutionLa3Ni2O7}%
  \BibitemOpen
  \bibfield  {author} {\bibinfo {author} {\bibfnamefont {Y.}~\bibnamefont {Meng}}, \bibinfo {author} {\bibfnamefont {Y.}~\bibnamefont {Yang}}, \bibinfo {author} {\bibfnamefont {H.}~\bibnamefont {Sun}}, \bibinfo {author} {\bibfnamefont {S.}~\bibnamefont {Zhang}}, \bibinfo {author} {\bibfnamefont {J.}~\bibnamefont {Luo}}, \bibinfo {author} {\bibfnamefont {M.}~\bibnamefont {Wang}}, \bibinfo {author} {\bibfnamefont {F.}~\bibnamefont {Hong}}, \bibinfo {author} {\bibfnamefont {X.}~\bibnamefont {Wang}},\ and\ \bibinfo {author} {\bibfnamefont {X.}~\bibnamefont {Yu}},\ }\href {https://doi.org/10.1038/s41467-024-54518-1} {\bibfield  {journal} {\bibinfo  {journal} {Nature Communications}\ }\textbf {\bibinfo {volume} {15}},\ \bibinfo {pages} {10408} (\bibinfo {year} {2024})}\BibitemShut {NoStop}%
\bibitem [{\citenamefont {Khasanov}\ \emph {et~al.}(2025)\citenamefont {Khasanov}, \citenamefont {Hicken}, \citenamefont {Gawryluk}, \citenamefont {Sorel}, \citenamefont {Bötzel}, \citenamefont {Lechermann}, \citenamefont {Eremin}, \citenamefont {Luetkens},\ and\ \citenamefont {Guguchia}}]{Khasanov2025}%
  \BibitemOpen
  \bibfield  {author} {\bibinfo {author} {\bibfnamefont {R.}~\bibnamefont {Khasanov}}, \bibinfo {author} {\bibfnamefont {T.~J.}\ \bibnamefont {Hicken}}, \bibinfo {author} {\bibfnamefont {D.~J.}\ \bibnamefont {Gawryluk}}, \bibinfo {author} {\bibfnamefont {L.~P.}\ \bibnamefont {Sorel}}, \bibinfo {author} {\bibfnamefont {S.}~\bibnamefont {Bötzel}}, \bibinfo {author} {\bibfnamefont {F.}~\bibnamefont {Lechermann}}, \bibinfo {author} {\bibfnamefont {I.~M.}\ \bibnamefont {Eremin}}, \bibinfo {author} {\bibfnamefont {H.}~\bibnamefont {Luetkens}},\ and\ \bibinfo {author} {\bibfnamefont {Z.}~\bibnamefont {Guguchia}},\ }\href {https://doi.org/10.1038/s41567-024-02754-z} {\bibfield  {journal} {\bibinfo  {journal} {Nature Physics}\ }\textbf {\bibinfo {volume} {21}},\ \bibinfo {pages} {430} (\bibinfo {year} {2025})}\BibitemShut {NoStop}%
\bibitem [{\citenamefont {Zhang}\ \emph {et~al.}(2023)\citenamefont {Zhang}, \citenamefont {Lin}, \citenamefont {Moreo},\ and\ \citenamefont {Dagotto}}]{PhysRevB.108.L180510}%
  \BibitemOpen
  \bibfield  {author} {\bibinfo {author} {\bibfnamefont {Y.}~\bibnamefont {Zhang}}, \bibinfo {author} {\bibfnamefont {L.-F.}\ \bibnamefont {Lin}}, \bibinfo {author} {\bibfnamefont {A.}~\bibnamefont {Moreo}},\ and\ \bibinfo {author} {\bibfnamefont {E.}~\bibnamefont {Dagotto}},\ }\href {https://doi.org/10.1103/PhysRevB.108.L180510} {\bibfield  {journal} {\bibinfo  {journal} {Phys. Rev. B}\ }\textbf {\bibinfo {volume} {108}},\ \bibinfo {pages} {L180510} (\bibinfo {year} {2023})}\BibitemShut {NoStop}%
\bibitem [{\citenamefont {Chen}\ \emph {et~al.}(2023)\citenamefont {Chen}, \citenamefont {Jiang}, \citenamefont {Li}, \citenamefont {Zhong},\ and\ \citenamefont {Lu}}]{chen2023criticalchargespininstabilities}%
  \BibitemOpen
  \bibfield  {author} {\bibinfo {author} {\bibfnamefont {X.}~\bibnamefont {Chen}}, \bibinfo {author} {\bibfnamefont {P.}~\bibnamefont {Jiang}}, \bibinfo {author} {\bibfnamefont {J.}~\bibnamefont {Li}}, \bibinfo {author} {\bibfnamefont {Z.}~\bibnamefont {Zhong}},\ and\ \bibinfo {author} {\bibfnamefont {Y.}~\bibnamefont {Lu}},\ }\href {https://arxiv.org/abs/2307.07154} {\bibinfo {title} {Critical charge and spin instabilities in superconducting $\rm{La}_{3}{Ni}_{2}{O}_{7}$}} (\bibinfo {year} {2023}),\ \Eprint {https://arxiv.org/abs/2307.07154} {arXiv:2307.07154 [cond-mat.supr-con]} \BibitemShut {NoStop}%
\bibitem [{\citenamefont {Lechermann}\ \emph {et~al.}(2023)\citenamefont {Lechermann}, \citenamefont {Gondolf}, \citenamefont {B\"otzel},\ and\ \citenamefont {Eremin}}]{PhysRevB.108.L201121}%
  \BibitemOpen
  \bibfield  {author} {\bibinfo {author} {\bibfnamefont {F.}~\bibnamefont {Lechermann}}, \bibinfo {author} {\bibfnamefont {J.}~\bibnamefont {Gondolf}}, \bibinfo {author} {\bibfnamefont {S.}~\bibnamefont {B\"otzel}},\ and\ \bibinfo {author} {\bibfnamefont {I.~M.}\ \bibnamefont {Eremin}},\ }\href {https://doi.org/10.1103/PhysRevB.108.L201121} {\bibfield  {journal} {\bibinfo  {journal} {Phys. Rev. B}\ }\textbf {\bibinfo {volume} {108}},\ \bibinfo {pages} {L201121} (\bibinfo {year} {2023})}\BibitemShut {NoStop}%
\bibitem [{\citenamefont {Gu}\ \emph {et~al.}(2023)\citenamefont {Gu}, \citenamefont {Le}, \citenamefont {Yang}, \citenamefont {Wu},\ and\ \citenamefont {Hu}}]{gu2023}%
  \BibitemOpen
  \bibfield  {author} {\bibinfo {author} {\bibfnamefont {Y.}~\bibnamefont {Gu}}, \bibinfo {author} {\bibfnamefont {C.}~\bibnamefont {Le}}, \bibinfo {author} {\bibfnamefont {Z.}~\bibnamefont {Yang}}, \bibinfo {author} {\bibfnamefont {X.}~\bibnamefont {Wu}},\ and\ \bibinfo {author} {\bibfnamefont {J.}~\bibnamefont {Hu}},\ }\href {https://arxiv.org/abs/2306.07275} {\bibinfo {title} {Effective model and pairing tendency in bilayer ni-based superconductor $\rm{La}_{3}{Ni}_{2}{O}_{7}$}} (\bibinfo {year} {2023}),\ \Eprint {https://arxiv.org/abs/2306.07275} {arXiv:2306.07275 [cond-mat.supr-con]} \BibitemShut {NoStop}%
\bibitem [{\citenamefont {Yang}\ \emph {et~al.}(2023{\natexlab{a}})\citenamefont {Yang}, \citenamefont {Wang},\ and\ \citenamefont {Wang}}]{PhysRevB.108.L140505}%
  \BibitemOpen
  \bibfield  {author} {\bibinfo {author} {\bibfnamefont {Q.-G.}\ \bibnamefont {Yang}}, \bibinfo {author} {\bibfnamefont {D.}~\bibnamefont {Wang}},\ and\ \bibinfo {author} {\bibfnamefont {Q.-H.}\ \bibnamefont {Wang}},\ }\href {https://doi.org/10.1103/PhysRevB.108.L140505} {\bibfield  {journal} {\bibinfo  {journal} {Phys. Rev. B}\ }\textbf {\bibinfo {volume} {108}},\ \bibinfo {pages} {L140505} (\bibinfo {year} {2023}{\natexlab{a}})}\BibitemShut {NoStop}%
\bibitem [{\citenamefont {Liu}\ \emph {et~al.}(2023)\citenamefont {Liu}, \citenamefont {Mei}, \citenamefont {Ye}, \citenamefont {Chen},\ and\ \citenamefont {Yang}}]{PhysRevLett.131.236002}%
  \BibitemOpen
  \bibfield  {author} {\bibinfo {author} {\bibfnamefont {Y.-B.}\ \bibnamefont {Liu}}, \bibinfo {author} {\bibfnamefont {J.-W.}\ \bibnamefont {Mei}}, \bibinfo {author} {\bibfnamefont {F.}~\bibnamefont {Ye}}, \bibinfo {author} {\bibfnamefont {W.-Q.}\ \bibnamefont {Chen}},\ and\ \bibinfo {author} {\bibfnamefont {F.}~\bibnamefont {Yang}},\ }\href {https://doi.org/10.1103/PhysRevLett.131.236002} {\bibfield  {journal} {\bibinfo  {journal} {Phys. Rev. Lett.}\ }\textbf {\bibinfo {volume} {131}},\ \bibinfo {pages} {236002} (\bibinfo {year} {2023})}\BibitemShut {NoStop}%
\bibitem [{\citenamefont {Yang}\ \emph {et~al.}(2023{\natexlab{b}})\citenamefont {Yang}, \citenamefont {Zhang},\ and\ \citenamefont {Zhang}}]{PhysRevB.108.L201108}%
  \BibitemOpen
  \bibfield  {author} {\bibinfo {author} {\bibfnamefont {Y.-f.}\ \bibnamefont {Yang}}, \bibinfo {author} {\bibfnamefont {G.-M.}\ \bibnamefont {Zhang}},\ and\ \bibinfo {author} {\bibfnamefont {F.-C.}\ \bibnamefont {Zhang}},\ }\href {https://doi.org/10.1103/PhysRevB.108.L201108} {\bibfield  {journal} {\bibinfo  {journal} {Phys. Rev. B}\ }\textbf {\bibinfo {volume} {108}},\ \bibinfo {pages} {L201108} (\bibinfo {year} {2023}{\natexlab{b}})}\BibitemShut {NoStop}%
\bibitem [{\citenamefont {Sakakibara}\ \emph {et~al.}(2024{\natexlab{b}})\citenamefont {Sakakibara}, \citenamefont {Kitamine}, \citenamefont {Ochi},\ and\ \citenamefont {Kuroki}}]{PhysRevLett.132.106002}%
  \BibitemOpen
  \bibfield  {author} {\bibinfo {author} {\bibfnamefont {H.}~\bibnamefont {Sakakibara}}, \bibinfo {author} {\bibfnamefont {N.}~\bibnamefont {Kitamine}}, \bibinfo {author} {\bibfnamefont {M.}~\bibnamefont {Ochi}},\ and\ \bibinfo {author} {\bibfnamefont {K.}~\bibnamefont {Kuroki}},\ }\href {https://doi.org/10.1103/PhysRevLett.132.106002} {\bibfield  {journal} {\bibinfo  {journal} {Phys. Rev. Lett.}\ }\textbf {\bibinfo {volume} {132}},\ \bibinfo {pages} {106002} (\bibinfo {year} {2024}{\natexlab{b}})}\BibitemShut {NoStop}%
\bibitem [{\citenamefont {Shen}\ \emph {et~al.}(2023)\citenamefont {Shen}, \citenamefont {Qin},\ and\ \citenamefont {Zhang}}]{Shen_2023}%
  \BibitemOpen
  \bibfield  {author} {\bibinfo {author} {\bibfnamefont {Y.}~\bibnamefont {Shen}}, \bibinfo {author} {\bibfnamefont {M.}~\bibnamefont {Qin}},\ and\ \bibinfo {author} {\bibfnamefont {G.-M.}\ \bibnamefont {Zhang}},\ }\href {https://doi.org/10.1088/0256-307X/40/12/127401} {\bibfield  {journal} {\bibinfo  {journal} {Chinese Physics Letters}\ }\textbf {\bibinfo {volume} {40}},\ \bibinfo {pages} {127401} (\bibinfo {year} {2023})}\BibitemShut {NoStop}%
\bibitem [{\citenamefont {Qu}\ \emph {et~al.}(2024)\citenamefont {Qu}, \citenamefont {Qu}, \citenamefont {Chen}, \citenamefont {Wu}, \citenamefont {Yang}, \citenamefont {Li},\ and\ \citenamefont {Su}}]{PhysRevLett.132.036502}%
  \BibitemOpen
  \bibfield  {author} {\bibinfo {author} {\bibfnamefont {X.-Z.}\ \bibnamefont {Qu}}, \bibinfo {author} {\bibfnamefont {D.-W.}\ \bibnamefont {Qu}}, \bibinfo {author} {\bibfnamefont {J.}~\bibnamefont {Chen}}, \bibinfo {author} {\bibfnamefont {C.}~\bibnamefont {Wu}}, \bibinfo {author} {\bibfnamefont {F.}~\bibnamefont {Yang}}, \bibinfo {author} {\bibfnamefont {W.}~\bibnamefont {Li}},\ and\ \bibinfo {author} {\bibfnamefont {G.}~\bibnamefont {Su}},\ }\href {https://doi.org/10.1103/PhysRevLett.132.036502} {\bibfield  {journal} {\bibinfo  {journal} {Phys. Rev. Lett.}\ }\textbf {\bibinfo {volume} {132}},\ \bibinfo {pages} {036502} (\bibinfo {year} {2024})}\BibitemShut {NoStop}%
\bibitem [{\citenamefont {Cao}\ and\ \citenamefont {Yang}(2024)}]{PhysRevB.109.L081105}%
  \BibitemOpen
  \bibfield  {author} {\bibinfo {author} {\bibfnamefont {Y.}~\bibnamefont {Cao}}\ and\ \bibinfo {author} {\bibfnamefont {Y.-f.}\ \bibnamefont {Yang}},\ }\href {https://doi.org/10.1103/PhysRevB.109.L081105} {\bibfield  {journal} {\bibinfo  {journal} {Phys. Rev. B}\ }\textbf {\bibinfo {volume} {109}},\ \bibinfo {pages} {L081105} (\bibinfo {year} {2024})}\BibitemShut {NoStop}%
\bibitem [{\citenamefont {Tian}\ \emph {et~al.}(2024)\citenamefont {Tian}, \citenamefont {Chen}, \citenamefont {Wang}, \citenamefont {He},\ and\ \citenamefont {Lu}}]{PhysRevB.109.165154}%
  \BibitemOpen
  \bibfield  {author} {\bibinfo {author} {\bibfnamefont {Y.-H.}\ \bibnamefont {Tian}}, \bibinfo {author} {\bibfnamefont {Y.}~\bibnamefont {Chen}}, \bibinfo {author} {\bibfnamefont {J.-M.}\ \bibnamefont {Wang}}, \bibinfo {author} {\bibfnamefont {R.-Q.}\ \bibnamefont {He}},\ and\ \bibinfo {author} {\bibfnamefont {Z.-Y.}\ \bibnamefont {Lu}},\ }\href {https://doi.org/10.1103/PhysRevB.109.165154} {\bibfield  {journal} {\bibinfo  {journal} {Phys. Rev. B}\ }\textbf {\bibinfo {volume} {109}},\ \bibinfo {pages} {165154} (\bibinfo {year} {2024})}\BibitemShut {NoStop}%
\bibitem [{\citenamefont {Lu}\ \emph {et~al.}(2023)\citenamefont {Lu}, \citenamefont {Li}, \citenamefont {Zeng}, \citenamefont {Hou}, \citenamefont {Wang}, \citenamefont {Yang},\ and\ \citenamefont {You}}]{lu2023superconductivitydopingsymmetricmass}%
  \BibitemOpen
  \bibfield  {author} {\bibinfo {author} {\bibfnamefont {D.-C.}\ \bibnamefont {Lu}}, \bibinfo {author} {\bibfnamefont {M.}~\bibnamefont {Li}}, \bibinfo {author} {\bibfnamefont {Z.-Y.}\ \bibnamefont {Zeng}}, \bibinfo {author} {\bibfnamefont {W.}~\bibnamefont {Hou}}, \bibinfo {author} {\bibfnamefont {J.}~\bibnamefont {Wang}}, \bibinfo {author} {\bibfnamefont {F.}~\bibnamefont {Yang}},\ and\ \bibinfo {author} {\bibfnamefont {Y.-Z.}\ \bibnamefont {You}},\ }\href {https://arxiv.org/abs/2308.11195} {\bibinfo {title} {Superconductivity from doping symmetric mass generation insulators: Application to $\rm{La}_{3}{Ni}_{2}{O}_{7}$ under pressure}} (\bibinfo {year} {2023}),\ \Eprint {https://arxiv.org/abs/2308.11195} {arXiv:2308.11195 [cond-mat.str-el]} \BibitemShut {NoStop}%
\bibitem [{\citenamefont {Huang}\ \emph {et~al.}(2023)\citenamefont {Huang}, \citenamefont {Wang},\ and\ \citenamefont {Zhou}}]{PhysRevB.108.174501}%
  \BibitemOpen
  \bibfield  {author} {\bibinfo {author} {\bibfnamefont {J.}~\bibnamefont {Huang}}, \bibinfo {author} {\bibfnamefont {Z.~D.}\ \bibnamefont {Wang}},\ and\ \bibinfo {author} {\bibfnamefont {T.}~\bibnamefont {Zhou}},\ }\href {https://doi.org/10.1103/PhysRevB.108.174501} {\bibfield  {journal} {\bibinfo  {journal} {Phys. Rev. B}\ }\textbf {\bibinfo {volume} {108}},\ \bibinfo {pages} {174501} (\bibinfo {year} {2023})}\BibitemShut {NoStop}%
\bibitem [{\citenamefont {Jiang}\ \emph {et~al.}(2024)\citenamefont {Jiang}, \citenamefont {Hou}, \citenamefont {Fan}, \citenamefont {Lang},\ and\ \citenamefont {Ku}}]{PhysRevLett.132.126503}%
  \BibitemOpen
  \bibfield  {author} {\bibinfo {author} {\bibfnamefont {R.}~\bibnamefont {Jiang}}, \bibinfo {author} {\bibfnamefont {J.}~\bibnamefont {Hou}}, \bibinfo {author} {\bibfnamefont {Z.}~\bibnamefont {Fan}}, \bibinfo {author} {\bibfnamefont {Z.-J.}\ \bibnamefont {Lang}},\ and\ \bibinfo {author} {\bibfnamefont {W.}~\bibnamefont {Ku}},\ }\href {https://doi.org/10.1103/PhysRevLett.132.126503} {\bibfield  {journal} {\bibinfo  {journal} {Phys. Rev. Lett.}\ }\textbf {\bibinfo {volume} {132}},\ \bibinfo {pages} {126503} (\bibinfo {year} {2024})}\BibitemShut {NoStop}%
\bibitem [{\citenamefont {Liao}\ \emph {et~al.}(2023)\citenamefont {Liao}, \citenamefont {Chen}, \citenamefont {Duan}, \citenamefont {Wang}, \citenamefont {Liu}, \citenamefont {Yu},\ and\ \citenamefont {Si}}]{PhysRevB.108.214522}%
  \BibitemOpen
  \bibfield  {author} {\bibinfo {author} {\bibfnamefont {Z.}~\bibnamefont {Liao}}, \bibinfo {author} {\bibfnamefont {L.}~\bibnamefont {Chen}}, \bibinfo {author} {\bibfnamefont {G.}~\bibnamefont {Duan}}, \bibinfo {author} {\bibfnamefont {Y.}~\bibnamefont {Wang}}, \bibinfo {author} {\bibfnamefont {C.}~\bibnamefont {Liu}}, \bibinfo {author} {\bibfnamefont {R.}~\bibnamefont {Yu}},\ and\ \bibinfo {author} {\bibfnamefont {Q.}~\bibnamefont {Si}},\ }\href {https://doi.org/10.1103/PhysRevB.108.214522} {\bibfield  {journal} {\bibinfo  {journal} {Phys. Rev. B}\ }\textbf {\bibinfo {volume} {108}},\ \bibinfo {pages} {214522} (\bibinfo {year} {2023})}\BibitemShut {NoStop}%
\bibitem [{\citenamefont {Lu}\ \emph {et~al.}(2024)\citenamefont {Lu}, \citenamefont {Pan}, \citenamefont {Yang},\ and\ \citenamefont {Wu}}]{PhysRevLett.132.146002}%
  \BibitemOpen
  \bibfield  {author} {\bibinfo {author} {\bibfnamefont {C.}~\bibnamefont {Lu}}, \bibinfo {author} {\bibfnamefont {Z.}~\bibnamefont {Pan}}, \bibinfo {author} {\bibfnamefont {F.}~\bibnamefont {Yang}},\ and\ \bibinfo {author} {\bibfnamefont {C.}~\bibnamefont {Wu}},\ }\href {https://doi.org/10.1103/PhysRevLett.132.146002} {\bibfield  {journal} {\bibinfo  {journal} {Phys. Rev. Lett.}\ }\textbf {\bibinfo {volume} {132}},\ \bibinfo {pages} {146002} (\bibinfo {year} {2024})}\BibitemShut {NoStop}%
\bibitem [{\citenamefont {Oh}\ and\ \citenamefont {Zhang}(2023)}]{PhysRevB.108.174511}%
  \BibitemOpen
  \bibfield  {author} {\bibinfo {author} {\bibfnamefont {H.}~\bibnamefont {Oh}}\ and\ \bibinfo {author} {\bibfnamefont {Y.-H.}\ \bibnamefont {Zhang}},\ }\href {https://doi.org/10.1103/PhysRevB.108.174511} {\bibfield  {journal} {\bibinfo  {journal} {Phys. Rev. B}\ }\textbf {\bibinfo {volume} {108}},\ \bibinfo {pages} {174511} (\bibinfo {year} {2023})}\BibitemShut {NoStop}%
\bibitem [{\citenamefont {Qin}\ and\ \citenamefont {Yang}(2023)}]{PhysRevB.108.L140504}%
  \BibitemOpen
  \bibfield  {author} {\bibinfo {author} {\bibfnamefont {Q.}~\bibnamefont {Qin}}\ and\ \bibinfo {author} {\bibfnamefont {Y.-f.}\ \bibnamefont {Yang}},\ }\href {https://doi.org/10.1103/PhysRevB.108.L140504} {\bibfield  {journal} {\bibinfo  {journal} {Phys. Rev. B}\ }\textbf {\bibinfo {volume} {108}},\ \bibinfo {pages} {L140504} (\bibinfo {year} {2023})}\BibitemShut {NoStop}%
\bibitem [{\citenamefont {Shilenko}\ and\ \citenamefont {Leonov}(2023)}]{PhysRevB.108.125105}%
  \BibitemOpen
  \bibfield  {author} {\bibinfo {author} {\bibfnamefont {D.~A.}\ \bibnamefont {Shilenko}}\ and\ \bibinfo {author} {\bibfnamefont {I.~V.}\ \bibnamefont {Leonov}},\ }\href {https://doi.org/10.1103/PhysRevB.108.125105} {\bibfield  {journal} {\bibinfo  {journal} {Phys. Rev. B}\ }\textbf {\bibinfo {volume} {108}},\ \bibinfo {pages} {125105} (\bibinfo {year} {2023})}\BibitemShut {NoStop}%
\bibitem [{\citenamefont {Yi}\ \emph {et~al.}(2024)\citenamefont {Yi}, \citenamefont {Meng}, \citenamefont {Li}, \citenamefont {Liao}, \citenamefont {Li}, \citenamefont {You}, \citenamefont {Gu},\ and\ \citenamefont {Su}}]{PhysRevB.110.L140508}%
  \BibitemOpen
  \bibfield  {author} {\bibinfo {author} {\bibfnamefont {X.-W.}\ \bibnamefont {Yi}}, \bibinfo {author} {\bibfnamefont {Y.}~\bibnamefont {Meng}}, \bibinfo {author} {\bibfnamefont {J.-W.}\ \bibnamefont {Li}}, \bibinfo {author} {\bibfnamefont {Z.-W.}\ \bibnamefont {Liao}}, \bibinfo {author} {\bibfnamefont {W.}~\bibnamefont {Li}}, \bibinfo {author} {\bibfnamefont {J.-Y.}\ \bibnamefont {You}}, \bibinfo {author} {\bibfnamefont {B.}~\bibnamefont {Gu}},\ and\ \bibinfo {author} {\bibfnamefont {G.}~\bibnamefont {Su}},\ }\href {https://doi.org/10.1103/PhysRevB.110.L140508} {\bibfield  {journal} {\bibinfo  {journal} {Phys. Rev. B}\ }\textbf {\bibinfo {volume} {110}},\ \bibinfo {pages} {L140508} (\bibinfo {year} {2024})}\BibitemShut {NoStop}%
\bibitem [{\citenamefont {Kaneko}\ \emph {et~al.}(2024)\citenamefont {Kaneko}, \citenamefont {Sakakibara}, \citenamefont {Ochi},\ and\ \citenamefont {Kuroki}}]{PhysRevB.109.045154}%
  \BibitemOpen
  \bibfield  {author} {\bibinfo {author} {\bibfnamefont {T.}~\bibnamefont {Kaneko}}, \bibinfo {author} {\bibfnamefont {H.}~\bibnamefont {Sakakibara}}, \bibinfo {author} {\bibfnamefont {M.}~\bibnamefont {Ochi}},\ and\ \bibinfo {author} {\bibfnamefont {K.}~\bibnamefont {Kuroki}},\ }\href {https://doi.org/10.1103/PhysRevB.109.045154} {\bibfield  {journal} {\bibinfo  {journal} {Phys. Rev. B}\ }\textbf {\bibinfo {volume} {109}},\ \bibinfo {pages} {045154} (\bibinfo {year} {2024})}\BibitemShut {NoStop}%
\bibitem [{\citenamefont {Qin}\ \emph {et~al.}(2024)\citenamefont {Qin}, \citenamefont {Foyevtsova}, \citenamefont {Si}, \citenamefont {Sawatzky},\ and\ \citenamefont {Jiang}}]{qin2024intertwinedchargespindensity}%
  \BibitemOpen
  \bibfield  {author} {\bibinfo {author} {\bibfnamefont {C.}~\bibnamefont {Qin}}, \bibinfo {author} {\bibfnamefont {K.}~\bibnamefont {Foyevtsova}}, \bibinfo {author} {\bibfnamefont {L.}~\bibnamefont {Si}}, \bibinfo {author} {\bibfnamefont {G.~A.}\ \bibnamefont {Sawatzky}},\ and\ \bibinfo {author} {\bibfnamefont {M.}~\bibnamefont {Jiang}},\ }\href {https://arxiv.org/abs/2410.15649} {\bibinfo {title} {Intertwined charge and spin density wave state of $\rm{La}_{3}{Ni}_{2}{O}_{7}$}} (\bibinfo {year} {2024}),\ \Eprint {https://arxiv.org/abs/2410.15649} {arXiv:2410.15649 [cond-mat.supr-con]} \BibitemShut {NoStop}%
\bibitem [{\citenamefont {Zhang}\ \emph {et~al.}(2024{\natexlab{c}})\citenamefont {Zhang}, \citenamefont {Bai}, \citenamefont {Kong}, \citenamefont {Wu}, \citenamefont {Xing},\ and\ \citenamefont {Xu}}]{zhang2024dopingevolutionnormalstate}%
  \BibitemOpen
  \bibfield  {author} {\bibinfo {author} {\bibfnamefont {H.-Y.}\ \bibnamefont {Zhang}}, \bibinfo {author} {\bibfnamefont {Y.-J.}\ \bibnamefont {Bai}}, \bibinfo {author} {\bibfnamefont {F.-J.}\ \bibnamefont {Kong}}, \bibinfo {author} {\bibfnamefont {X.-Q.}\ \bibnamefont {Wu}}, \bibinfo {author} {\bibfnamefont {Y.-H.}\ \bibnamefont {Xing}},\ and\ \bibinfo {author} {\bibfnamefont {N.}~\bibnamefont {Xu}},\ }\href {https://arxiv.org/abs/2408.03763} {\bibinfo {title} {Doping evolution of the normal state magnetic excitations in pressurized $\rm{La}_{3}{Ni}_{2}{O}_{7}$}} (\bibinfo {year} {2024}{\natexlab{c}}),\ \Eprint {https://arxiv.org/abs/2408.03763} {arXiv:2408.03763 [cond-mat.supr-con]} \BibitemShut {NoStop}%
\bibitem [{\citenamefont {Zhang}\ \emph {et~al.}(2024{\natexlab{d}})\citenamefont {Zhang}, \citenamefont {Lin}, \citenamefont {Moreo}, \citenamefont {Maier},\ and\ \citenamefont {Dagotto}}]{Zhang2024stripe}%
  \BibitemOpen
  \bibfield  {author} {\bibinfo {author} {\bibfnamefont {Y.}~\bibnamefont {Zhang}}, \bibinfo {author} {\bibfnamefont {L.-F.}\ \bibnamefont {Lin}}, \bibinfo {author} {\bibfnamefont {A.}~\bibnamefont {Moreo}}, \bibinfo {author} {\bibfnamefont {T.~A.}\ \bibnamefont {Maier}},\ and\ \bibinfo {author} {\bibfnamefont {E.}~\bibnamefont {Dagotto}},\ }\bibfield  {journal} {\bibinfo  {journal} {Nature Communications}\ }\textbf {\bibinfo {volume} {15}},\ \href {https://doi.org/10.1038/s41467-024-46622-z} {10.1038/s41467-024-46622-z} (\bibinfo {year} {2024}{\natexlab{d}})\BibitemShut {NoStop}%
\bibitem [{\citenamefont {LaBollita}\ \emph {et~al.}(2024)\citenamefont {LaBollita}, \citenamefont {Pardo}, \citenamefont {Norman},\ and\ \citenamefont {Botana}}]{labollita2024assessingformationspincharge}%
  \BibitemOpen
  \bibfield  {author} {\bibinfo {author} {\bibfnamefont {H.}~\bibnamefont {LaBollita}}, \bibinfo {author} {\bibfnamefont {V.}~\bibnamefont {Pardo}}, \bibinfo {author} {\bibfnamefont {M.~R.}\ \bibnamefont {Norman}},\ and\ \bibinfo {author} {\bibfnamefont {A.~S.}\ \bibnamefont {Botana}},\ }\href {https://arxiv.org/abs/2407.14409} {\bibinfo {title} {Assessing the formation of spin and charge stripes in $\rm{La}_{3}{Ni}_{2}{O}_{7}$ from first-principles}} (\bibinfo {year} {2024}),\ \Eprint {https://arxiv.org/abs/2407.14409} {arXiv:2407.14409 [cond-mat.str-el]} \BibitemShut {NoStop}%
\bibitem [{\citenamefont {Zhang}\ \emph {et~al.}(2024{\natexlab{e}})\citenamefont {Zhang}, \citenamefont {Xu},\ and\ \citenamefont {Xiang}}]{zhang2024emergentspinchargeorbitalordersuperconductor}%
  \BibitemOpen
  \bibfield  {author} {\bibinfo {author} {\bibfnamefont {B.}~\bibnamefont {Zhang}}, \bibinfo {author} {\bibfnamefont {C.}~\bibnamefont {Xu}},\ and\ \bibinfo {author} {\bibfnamefont {H.}~\bibnamefont {Xiang}},\ }\href {https://arxiv.org/abs/2407.18473} {\bibinfo {title} {Emergent spin-charge-orbital order in superconductor $\rm{La}_{3}{Ni}_{2}{O}_{7}$}} (\bibinfo {year} {2024}{\natexlab{e}}),\ \Eprint {https://arxiv.org/abs/2407.18473} {arXiv:2407.18473 [cond-mat.supr-con]} \BibitemShut {NoStop}%
\bibitem [{\citenamefont {Lin}\ \emph {et~al.}(2024{\natexlab{a}})\citenamefont {Lin}, \citenamefont {Zhang}, \citenamefont {Kaushal}, \citenamefont {Alvarez}, \citenamefont {Maier}, \citenamefont {Moreo},\ and\ \citenamefont {Dagotto}}]{lin2024magneticphasediagramtwoorbitaL}%
  \BibitemOpen
  \bibfield  {author} {\bibinfo {author} {\bibfnamefont {L.-F.}\ \bibnamefont {Lin}}, \bibinfo {author} {\bibfnamefont {Y.}~\bibnamefont {Zhang}}, \bibinfo {author} {\bibfnamefont {N.}~\bibnamefont {Kaushal}}, \bibinfo {author} {\bibfnamefont {G.}~\bibnamefont {Alvarez}}, \bibinfo {author} {\bibfnamefont {T.~A.}\ \bibnamefont {Maier}}, \bibinfo {author} {\bibfnamefont {A.}~\bibnamefont {Moreo}},\ and\ \bibinfo {author} {\bibfnamefont {E.}~\bibnamefont {Dagotto}},\ }\href {https://arxiv.org/abs/2408.05689} {\bibinfo {title} {Magnetic phase diagram of a two-orbital model for bilayer nickelates varying doping}} (\bibinfo {year} {2024}{\natexlab{a}}),\ \Eprint {https://arxiv.org/abs/2408.05689} {arXiv:2408.05689 [cond-mat.str-el]} \BibitemShut {NoStop}%
\bibitem [{\citenamefont {Lin}\ \emph {et~al.}(2024{\natexlab{b}})\citenamefont {Lin}, \citenamefont {Zhang}, \citenamefont {Kaushal}, \citenamefont {Alvarez}, \citenamefont {Maier}, \citenamefont {Moreo},\ and\ \citenamefont {Dagotto}}]{PhysRevB.110.195135}%
  \BibitemOpen
  \bibfield  {author} {\bibinfo {author} {\bibfnamefont {L.-F.}\ \bibnamefont {Lin}}, \bibinfo {author} {\bibfnamefont {Y.}~\bibnamefont {Zhang}}, \bibinfo {author} {\bibfnamefont {N.}~\bibnamefont {Kaushal}}, \bibinfo {author} {\bibfnamefont {G.}~\bibnamefont {Alvarez}}, \bibinfo {author} {\bibfnamefont {T.~A.}\ \bibnamefont {Maier}}, \bibinfo {author} {\bibfnamefont {A.}~\bibnamefont {Moreo}},\ and\ \bibinfo {author} {\bibfnamefont {E.}~\bibnamefont {Dagotto}},\ }\href {https://doi.org/10.1103/PhysRevB.110.195135} {\bibfield  {journal} {\bibinfo  {journal} {Phys. Rev. B}\ }\textbf {\bibinfo {volume} {110}},\ \bibinfo {pages} {195135} (\bibinfo {year} {2024}{\natexlab{b}})}\BibitemShut {NoStop}%
\bibitem [{\citenamefont {Fan}\ \emph {et~al.}(2024)\citenamefont {Fan}, \citenamefont {Zhang}, \citenamefont {Zhan}, \citenamefont {Lv}, \citenamefont {Jiang}, \citenamefont {Normand},\ and\ \citenamefont {Xiang}}]{PhysRevB.110.024514}%
  \BibitemOpen
  \bibfield  {author} {\bibinfo {author} {\bibfnamefont {Z.}~\bibnamefont {Fan}}, \bibinfo {author} {\bibfnamefont {J.-F.}\ \bibnamefont {Zhang}}, \bibinfo {author} {\bibfnamefont {B.}~\bibnamefont {Zhan}}, \bibinfo {author} {\bibfnamefont {D.}~\bibnamefont {Lv}}, \bibinfo {author} {\bibfnamefont {X.-Y.}\ \bibnamefont {Jiang}}, \bibinfo {author} {\bibfnamefont {B.}~\bibnamefont {Normand}},\ and\ \bibinfo {author} {\bibfnamefont {T.}~\bibnamefont {Xiang}},\ }\href {https://doi.org/10.1103/PhysRevB.110.024514} {\bibfield  {journal} {\bibinfo  {journal} {Phys. Rev. B}\ }\textbf {\bibinfo {volume} {110}},\ \bibinfo {pages} {024514} (\bibinfo {year} {2024})}\BibitemShut {NoStop}%
\bibitem [{\citenamefont {Schlömer}\ \emph {et~al.}(2024)\citenamefont {Schlömer}, \citenamefont {Schollwöck}, \citenamefont {Grusdt},\ and\ \citenamefont {Bohrdt}}]{schlömer2024superconductivitypressurizednickelatela3ni2o7}%
  \BibitemOpen
  \bibfield  {author} {\bibinfo {author} {\bibfnamefont {H.}~\bibnamefont {Schlömer}}, \bibinfo {author} {\bibfnamefont {U.}~\bibnamefont {Schollwöck}}, \bibinfo {author} {\bibfnamefont {F.}~\bibnamefont {Grusdt}},\ and\ \bibinfo {author} {\bibfnamefont {A.}~\bibnamefont {Bohrdt}},\ }\href {https://arxiv.org/abs/2311.03349} {\bibinfo {title} {Superconductivity in the pressurized nickelate $\rm{La}_{3}{Ni}_{2}{O}_{7}$ in the vicinity of a bec-bcs crossover}} (\bibinfo {year} {2024}),\ \Eprint {https://arxiv.org/abs/2311.03349} {arXiv:2311.03349 [cond-mat.str-el]} \BibitemShut {NoStop}%
\bibitem [{\citenamefont {Zhang}\ \emph {et~al.}(2024{\natexlab{f}})\citenamefont {Zhang}, \citenamefont {Zhang}, \citenamefont {You},\ and\ \citenamefont {Weng}}]{PhysRevLett.133.126501}%
  \BibitemOpen
  \bibfield  {author} {\bibinfo {author} {\bibfnamefont {J.-X.}\ \bibnamefont {Zhang}}, \bibinfo {author} {\bibfnamefont {H.-K.}\ \bibnamefont {Zhang}}, \bibinfo {author} {\bibfnamefont {Y.-Z.}\ \bibnamefont {You}},\ and\ \bibinfo {author} {\bibfnamefont {Z.-Y.}\ \bibnamefont {Weng}},\ }\href {https://doi.org/10.1103/PhysRevLett.133.126501} {\bibfield  {journal} {\bibinfo  {journal} {Phys. Rev. Lett.}\ }\textbf {\bibinfo {volume} {133}},\ \bibinfo {pages} {126501} (\bibinfo {year} {2024}{\natexlab{f}})}\BibitemShut {NoStop}%
\bibitem [{\citenamefont {Zhang}\ and\ \citenamefont {Rice}(1988)}]{PhysRevB.37.3759}%
  \BibitemOpen
  \bibfield  {author} {\bibinfo {author} {\bibfnamefont {F.~C.}\ \bibnamefont {Zhang}}\ and\ \bibinfo {author} {\bibfnamefont {T.~M.}\ \bibnamefont {Rice}},\ }\href {https://doi.org/10.1103/PhysRevB.37.3759} {\bibfield  {journal} {\bibinfo  {journal} {Phys. Rev. B}\ }\textbf {\bibinfo {volume} {37}},\ \bibinfo {pages} {3759} (\bibinfo {year} {1988})}\BibitemShut {NoStop}%
\bibitem [{\citenamefont {Zhang}\ \emph {et~al.}(2003)\citenamefont {Zhang}, \citenamefont {Gros}, \citenamefont {Rice},\ and\ \citenamefont {Shiba}}]{article2}%
  \BibitemOpen
  \bibfield  {author} {\bibinfo {author} {\bibfnamefont {F.}~\bibnamefont {Zhang}}, \bibinfo {author} {\bibfnamefont {C.}~\bibnamefont {Gros}}, \bibinfo {author} {\bibfnamefont {T.}~\bibnamefont {Rice}},\ and\ \bibinfo {author} {\bibfnamefont {H.}~\bibnamefont {Shiba}},\ }\href {https://doi.org/10.1088/0953-2048/1/1/009} {\bibfield  {journal} {\bibinfo  {journal} {Superconductor Science and Technology}\ }\textbf {\bibinfo {volume} {1}} (\bibinfo {year} {2003})}\BibitemShut {NoStop}%
\bibitem [{\citenamefont {Ogata}\ and\ \citenamefont {Himeda}(2003)}]{article}%
  \BibitemOpen
  \bibfield  {author} {\bibinfo {author} {\bibfnamefont {M.}~\bibnamefont {Ogata}}\ and\ \bibinfo {author} {\bibfnamefont {A.}~\bibnamefont {Himeda}},\ }\href {https://doi.org/10.1143/JPSJ.72.374} {\bibfield  {journal} {\bibinfo  {journal} {Journal of The Physical Society of Japan - J PHYS SOC JPN}\ }\textbf {\bibinfo {volume} {72}},\ \bibinfo {pages} {374} (\bibinfo {year} {2003})}\BibitemShut {NoStop}%
\bibitem [{\citenamefont {Himeda}\ and\ \citenamefont {Ogata}(1999)}]{PhysRevB.60.R9935}%
  \BibitemOpen
  \bibfield  {author} {\bibinfo {author} {\bibfnamefont {A.}~\bibnamefont {Himeda}}\ and\ \bibinfo {author} {\bibfnamefont {M.}~\bibnamefont {Ogata}},\ }\href {https://doi.org/10.1103/PhysRevB.60.R9935} {\bibfield  {journal} {\bibinfo  {journal} {Phys. Rev. B}\ }\textbf {\bibinfo {volume} {60}},\ \bibinfo {pages} {R9935} (\bibinfo {year} {1999})}\BibitemShut {NoStop}%
\bibitem [{\citenamefont {Yang}\ \emph {et~al.}(2009)\citenamefont {Yang}, \citenamefont {Chen}, \citenamefont {Rice}, \citenamefont {Sigrist},\ and\ \citenamefont {Zhang}}]{Yang2009}%
  \BibitemOpen
  \bibfield  {author} {\bibinfo {author} {\bibfnamefont {K.-Y.}\ \bibnamefont {Yang}}, \bibinfo {author} {\bibfnamefont {W.~Q.}\ \bibnamefont {Chen}}, \bibinfo {author} {\bibfnamefont {T.~M.}\ \bibnamefont {Rice}}, \bibinfo {author} {\bibfnamefont {M.}~\bibnamefont {Sigrist}},\ and\ \bibinfo {author} {\bibfnamefont {F.-C.}\ \bibnamefont {Zhang}},\ }\href {https://doi.org/10.1088/1367-2630/11/5/055053} {\bibfield  {journal} {\bibinfo  {journal} {New Journal of Physics}\ }\textbf {\bibinfo {volume} {11}},\ \bibinfo {pages} {055053} (\bibinfo {year} {2009})}\BibitemShut {NoStop}%
\bibitem [{\citenamefont {Chou}\ and\ \citenamefont {Lee}(2012)}]{PhysRevB.85.104511}%
  \BibitemOpen
  \bibfield  {author} {\bibinfo {author} {\bibfnamefont {C.-P.}\ \bibnamefont {Chou}}\ and\ \bibinfo {author} {\bibfnamefont {T.-K.}\ \bibnamefont {Lee}},\ }\href {https://doi.org/10.1103/PhysRevB.85.104511} {\bibfield  {journal} {\bibinfo  {journal} {Phys. Rev. B}\ }\textbf {\bibinfo {volume} {85}},\ \bibinfo {pages} {104511} (\bibinfo {year} {2012})}\BibitemShut {NoStop}%
\bibitem [{\citenamefont {Tu}\ and\ \citenamefont {Lee}(2016)}]{Tu2016}%
  \BibitemOpen
  \bibfield  {author} {\bibinfo {author} {\bibfnamefont {W.-L.}\ \bibnamefont {Tu}}\ and\ \bibinfo {author} {\bibfnamefont {T.-K.}\ \bibnamefont {Lee}},\ }\href {https://doi.org/10.1038/srep18675} {\bibfield  {journal} {\bibinfo  {journal} {Scientific Reports}\ }\textbf {\bibinfo {volume} {6}},\ \bibinfo {pages} {18675} (\bibinfo {year} {2016})}\BibitemShut {NoStop}%
\bibitem [{\citenamefont {{Yang}}\ \emph {et~al.}(2024)\citenamefont {{Yang}}, \citenamefont {{Sun}}, \citenamefont {{Hu}}, \citenamefont {{Xie}}, \citenamefont {{Miao}}, \citenamefont {{Luo}}, \citenamefont {{Chen}}, \citenamefont {{Liang}}, \citenamefont {{Zhu}}, \citenamefont {{Qu}}, \citenamefont {{Chen}}, \citenamefont {{Huo}}, \citenamefont {{Huang}}, \citenamefont {{Zhang}}, \citenamefont {{Zhang}}, \citenamefont {{Yang}}, \citenamefont {{Wang}}, \citenamefont {{Peng}}, \citenamefont {{Mao}}, \citenamefont {{Liu}}, \citenamefont {{Xu}}, \citenamefont {{Qian}}, \citenamefont {{Yao}}, \citenamefont {{Wang}}, \citenamefont {{Zhao}},\ and\ \citenamefont {{Zhou}}}]{Yang2024}%
  \BibitemOpen
  \bibfield  {author} {\bibinfo {author} {\bibfnamefont {J.}~\bibnamefont {{Yang}}}, \bibinfo {author} {\bibfnamefont {H.}~\bibnamefont {{Sun}}}, \bibinfo {author} {\bibfnamefont {X.}~\bibnamefont {{Hu}}}, \bibinfo {author} {\bibfnamefont {Y.}~\bibnamefont {{Xie}}}, \bibinfo {author} {\bibfnamefont {T.}~\bibnamefont {{Miao}}}, \bibinfo {author} {\bibfnamefont {H.}~\bibnamefont {{Luo}}}, \bibinfo {author} {\bibfnamefont {H.}~\bibnamefont {{Chen}}}, \bibinfo {author} {\bibfnamefont {B.}~\bibnamefont {{Liang}}}, \bibinfo {author} {\bibfnamefont {W.}~\bibnamefont {{Zhu}}}, \bibinfo {author} {\bibfnamefont {G.}~\bibnamefont {{Qu}}}, \bibinfo {author} {\bibfnamefont {C.-Q.}\ \bibnamefont {{Chen}}}, \bibinfo {author} {\bibfnamefont {M.}~\bibnamefont {{Huo}}}, \bibinfo {author} {\bibfnamefont {Y.}~\bibnamefont {{Huang}}}, \bibinfo {author} {\bibfnamefont {S.}~\bibnamefont {{Zhang}}}, \bibinfo {author} {\bibfnamefont {F.}~\bibnamefont {{Zhang}}}, \bibinfo {author} {\bibfnamefont {F.}~\bibnamefont {{Yang}}}, \bibinfo
  {author} {\bibfnamefont {Z.}~\bibnamefont {{Wang}}}, \bibinfo {author} {\bibfnamefont {Q.}~\bibnamefont {{Peng}}}, \bibinfo {author} {\bibfnamefont {H.}~\bibnamefont {{Mao}}}, \bibinfo {author} {\bibfnamefont {G.}~\bibnamefont {{Liu}}}, \bibinfo {author} {\bibfnamefont {Z.}~\bibnamefont {{Xu}}}, \bibinfo {author} {\bibfnamefont {T.}~\bibnamefont {{Qian}}}, \bibinfo {author} {\bibfnamefont {D.-X.}\ \bibnamefont {{Yao}}}, \bibinfo {author} {\bibfnamefont {M.}~\bibnamefont {{Wang}}}, \bibinfo {author} {\bibfnamefont {L.}~\bibnamefont {{Zhao}}},\ and\ \bibinfo {author} {\bibfnamefont {X.~J.}\ \bibnamefont {{Zhou}}},\ }\href {https://doi.org/10.1038/s41467-024-48701-7} {\bibfield  {journal} {\bibinfo  {journal} {Nature Communications}\ }\textbf {\bibinfo {volume} {15}},\ \bibinfo {pages} {4373} (\bibinfo {year} {2024})}\BibitemShut {NoStop}%
\bibitem [{\citenamefont {Luo}\ \emph {et~al.}(2024)\citenamefont {Luo}, \citenamefont {Lv}, \citenamefont {Wang}, \citenamefont {Wú},\ and\ \citenamefont {Yao}}]{Luo2024}%
  \BibitemOpen
  \bibfield  {author} {\bibinfo {author} {\bibfnamefont {Z.}~\bibnamefont {Luo}}, \bibinfo {author} {\bibfnamefont {B.}~\bibnamefont {Lv}}, \bibinfo {author} {\bibfnamefont {M.}~\bibnamefont {Wang}}, \bibinfo {author} {\bibfnamefont {W.}~\bibnamefont {Wú}},\ and\ \bibinfo {author} {\bibfnamefont {D.-X.}\ \bibnamefont {Yao}},\ }\href {https://doi.org/10.1038/s41535-024-00668-w} {\bibfield  {journal} {\bibinfo  {journal} {npj Quantum Materials}\ }\textbf {\bibinfo {volume} {9}},\ \bibinfo {pages} {61} (\bibinfo {year} {2024})}\BibitemShut {NoStop}%
\bibitem [{\citenamefont {Sigrist}\ \emph {et~al.}(1994)\citenamefont {Sigrist}, \citenamefont {Rice},\ and\ \citenamefont {Zhang}}]{PhysRevB.49.12058}%
  \BibitemOpen
  \bibfield  {author} {\bibinfo {author} {\bibfnamefont {M.}~\bibnamefont {Sigrist}}, \bibinfo {author} {\bibfnamefont {T.~M.}\ \bibnamefont {Rice}},\ and\ \bibinfo {author} {\bibfnamefont {F.~C.}\ \bibnamefont {Zhang}},\ }\href {https://doi.org/10.1103/PhysRevB.49.12058} {\bibfield  {journal} {\bibinfo  {journal} {Phys. Rev. B}\ }\textbf {\bibinfo {volume} {49}},\ \bibinfo {pages} {12058} (\bibinfo {year} {1994})}\BibitemShut {NoStop}%
\bibitem [{\citenamefont {Hou}\ \emph {et~al.}(2019{\natexlab{a}})\citenamefont {Hou}, \citenamefont {Lee}, \citenamefont {Guo}, \citenamefont {Lou},\ and\ \citenamefont {Chen}}]{Hou_2019}%
  \BibitemOpen
  \bibfield  {author} {\bibinfo {author} {\bibfnamefont {J.}~\bibnamefont {Hou}}, \bibinfo {author} {\bibfnamefont {T.-K.}\ \bibnamefont {Lee}}, \bibinfo {author} {\bibfnamefont {Y.}~\bibnamefont {Guo}}, \bibinfo {author} {\bibfnamefont {J.}~\bibnamefont {Lou}},\ and\ \bibinfo {author} {\bibfnamefont {Y.}~\bibnamefont {Chen}},\ }\href {https://doi.org/10.1088/1367-2630/ab5603} {\bibfield  {journal} {\bibinfo  {journal} {New Journal of Physics}\ }\textbf {\bibinfo {volume} {21}},\ \bibinfo {pages} {113047} (\bibinfo {year} {2019}{\natexlab{a}})}\BibitemShut {NoStop}%
\bibitem [{\citenamefont {Hou}\ \emph {et~al.}(2019{\natexlab{b}})\citenamefont {Hou}, \citenamefont {Lee}, \citenamefont {Lou},\ and\ \citenamefont {Chen}}]{PhysRevB.99.094510}%
  \BibitemOpen
  \bibfield  {author} {\bibinfo {author} {\bibfnamefont {J.}~\bibnamefont {Hou}}, \bibinfo {author} {\bibfnamefont {T.-K.}\ \bibnamefont {Lee}}, \bibinfo {author} {\bibfnamefont {J.}~\bibnamefont {Lou}},\ and\ \bibinfo {author} {\bibfnamefont {Y.}~\bibnamefont {Chen}},\ }\href {https://doi.org/10.1103/PhysRevB.99.094510} {\bibfield  {journal} {\bibinfo  {journal} {Phys. Rev. B}\ }\textbf {\bibinfo {volume} {99}},\ \bibinfo {pages} {094510} (\bibinfo {year} {2019}{\natexlab{b}})}\BibitemShut {NoStop}%
\bibitem [{\citenamefont {Dagotto}\ \emph {et~al.}(1992)\citenamefont {Dagotto}, \citenamefont {Riera},\ and\ \citenamefont {Scalapino}}]{PhysRevB.45.5744}%
  \BibitemOpen
  \bibfield  {author} {\bibinfo {author} {\bibfnamefont {E.}~\bibnamefont {Dagotto}}, \bibinfo {author} {\bibfnamefont {J.}~\bibnamefont {Riera}},\ and\ \bibinfo {author} {\bibfnamefont {D.}~\bibnamefont {Scalapino}},\ }\href {https://doi.org/10.1103/PhysRevB.45.5744} {\bibfield  {journal} {\bibinfo  {journal} {Phys. Rev. B}\ }\textbf {\bibinfo {volume} {45}},\ \bibinfo {pages} {5744} (\bibinfo {year} {1992})}\BibitemShut {NoStop}%
\bibitem [{\citenamefont {Wang}\ \emph {et~al.}(2024{\natexlab{b}})\citenamefont {Wang}, \citenamefont {Wen}, \citenamefont {Wu}, \citenamefont {Yao},\ and\ \citenamefont {Xiang}}]{cpl417077402}%
  \BibitemOpen
  \bibfield  {author} {\bibinfo {author} {\bibfnamefont {M.}~\bibnamefont {Wang}}, \bibinfo {author} {\bibfnamefont {H.-H.}\ \bibnamefont {Wen}}, \bibinfo {author} {\bibfnamefont {T.}~\bibnamefont {Wu}}, \bibinfo {author} {\bibfnamefont {D.-X.}\ \bibnamefont {Yao}},\ and\ \bibinfo {author} {\bibfnamefont {T.}~\bibnamefont {Xiang}},\ }\href {https://doi.org/10.1088/0256-307X/41/7/077402} {\bibfield  {journal} {\bibinfo  {journal} {Chin. Phys. Lett.}\ }\textbf {\bibinfo {volume} {41}} (\bibinfo {year} {2024}{\natexlab{b}})}\BibitemShut {NoStop}%
\bibitem [{\citenamefont {Wú}\ \emph {et~al.}(2024)\citenamefont {Wú}, \citenamefont {Luo}, \citenamefont {Yao},\ and\ \citenamefont {Wang}}]{Wu2024}%
  \BibitemOpen
  \bibfield  {author} {\bibinfo {author} {\bibfnamefont {W.}~\bibnamefont {Wú}}, \bibinfo {author} {\bibfnamefont {Z.}~\bibnamefont {Luo}}, \bibinfo {author} {\bibfnamefont {D.-X.}\ \bibnamefont {Yao}},\ and\ \bibinfo {author} {\bibfnamefont {M.}~\bibnamefont {Wang}},\ }\href {https://doi.org/10.1007/s11433-023-2300-4} {\bibfield  {journal} {\bibinfo  {journal} {Sci. China Phys. Mech. Astron.}\ }\textbf {\bibinfo {volume} {67}},\ \bibinfo {pages} {117402} (\bibinfo {year} {2024})}\BibitemShut {NoStop}%
\end{thebibliography}%
\end{document}